%% file: manuscript.tex
\documentclass[manuscript=article,journal=jpccck,layout=twocolumn]{achemso}

\input{head_packages}

\input{head_newcommands}


\input{head_newcommands_basic}

\input{head_newcommands_imp}

\input{head_modular}

\author{Fabian Single}
\affiliation[German Aerospace Center]{German Aerospace Center (DLR), Institute of Engineering Thermodynamics, Pfaffenwaldring 38-40, 70569 Stuttgart, Germany}
\alsoaffiliation[Helmholtz Institute Ulm]{Helmholtz Institute Ulm (HIU), Helmholtzstra\ss e, 11, 89081 Ulm, Germany}
\author{Birger Horstmann}
\alsoaffiliation[German Aerospace Center]{German Aerospace Center (DLR), Institute of Engineering Thermodynamics, Pfaffenwaldring 38-40, 70569 Stuttgart, Germany}
\alsoaffiliation[Helmholtz Institute Ulm]{Helmholtz Institute Ulm (HIU), Helmholtzstra\ss e, 11, 89081 Ulm, Germany}
\affiliation[University of Ulm]{University of Ulm, Albert-Einstein-Allee 47, 89081 Ulm, Germany}
\email{birger.horstmann@dlr.de}
\author{Arnulf Latz}
\alsoaffiliation[German Aerospace Center]{German Aerospace Center (DLR), Institute of Engineering Thermodynamics, Pfaffenwaldring 38-40, 70569 Stuttgart, Germany}
\alsoaffiliation[Helmholtz Institute Ulm]{Helmholtz Institute Ulm (HIU), Helmholtzstra\ss e, 11, 89081 Ulm, Germany}
\affiliation[University of Ulm]{University of Ulm, Albert-Einstein-Allee 47, 89081 Ulm, Germany}
\email{arnulf.latz@dlr.de}

\title{\titletext}

\begin{document}

\begin{abstract}
In this article, we derive and discuss a physics-based model for impedance spectroscopy of lithium batteries. Our model for electrochemical cells with planar electrodes takes into account the solid-electrolyte interphase (SEI) as porous surface film.
We present two improvements over standard impedance models. Firstly, our model is based on a consistent description of lithium transport through electrolyte and SEI. We use well-defined transport parameters, e.g., transference numbers, and consider convection of the center-of-mass. Secondly, we solve our model equations analytically and state the full transport parameter dependence of the impedance signals.
Our consistent model results in an analytic expression for the cell impedance including bulk and surface processes. The impedance signals due to concentration polarizations highlight the importance of electrolyte convection in concentrated electrolytes.
We simplify our expression for the complex impedance and compare it to common equivalent circuit models. Such simplified models are good approximations in concise parameter ranges.
Finally, we compare our model with experiments of lithium metal electrodes and find large transference numbers for lithium ions. This analysis reveals that lithium-ion transport through the SEI has solid electrolyte character.
\end{abstract}

\maketitle

\fsection[sec:introduction]{Introduction}
Impedance spectroscopy is an essential tool for the characterization of electrochemical devices.
This method gives insight into phenomena that are otherwise difficult to access.
Its non-destructive nature makes it especially suitable for monitoring delicate surface films such as the solid electrolyte interphase (SEI)\cite{Fauteux1985,Lee1988,Lu2014,Steinhauer2017a} in lithium-ion batteries.

Interpretation of impedance measurements requires a modeling approach.
Today, equivalent circuit models remain the most prominent model type for this purpose \cite{Levi1997}.
However, such models often mask the way some parameters influence the impedance.
Physics-based models include these dependencies at the cost of an increased modeling effort.
Numerous comprehensive models exist and describe a diverse amount of electrochemical processes and systems.
These include cell level models of standard lithium-ion batteries\cite{Doyle1993a,Danner2016}, Li-sulfur batteries\cite{Fronczek2013,Danner2015}, metal-air batteries\cite{Schmitt2019,C9TA01190K,Clark2017b}, and fuel cells\cite{Jahnke2017}.
Other models focus on selected electrochemical processes of interest, such as membranes \cite{Moya2016}, interface reactions \cite{Huang2012,Luck2019}, electrochemical double layers \cite{Braun2015,Hoffmann2018}, and growth of surface layers \cite{Single2016, Horstmann2013b}. 
Such models accurately capture reactions and transport in the complex geometry and morphology of the corresponding system.
They are also used for impedance calculations by taking into account interface\remove{ reactions and} capacities\cite{Legrand2014}.
Then, one can calculate the cell impedance with a single voltage step simulation\cite{Bessler2007}.

Several models discuss the impedance of lithium-ion batteries.
Most of these models go to great lengths to describe the porous electrode and the frequency dependent response of single electrode particles.
This is described within the framework of 1+1D Newman models \cite{Levi2004,Huang2006}.
Advanced models consider a particle size distribution\cite{Meyers2000} and anisotropic particles\cite{Song2012}.
More recent publications also discuss the distribution of relaxation time\ins{s} \cite{Schmidt2011} and consider higher harmonics \cite{Murbach2017}. %P2D imp. higher harmonics

Most of the impedance models cited above are either semi-analytic or fully numeric.
However, only exact analytic models allow the use of impedance spectroscopy to determine parameters and physical quantities, e.g., diffusion coefficient, Tafel slope, and double layer capacitance.
This is the added value of exact analytic results as demonstrated by  Kulikovsky et al. with multiple impedance models for fuel cells \cite{Kulikovsky2012,Kulikovsky2013,Kulikovsky2015}.

In our impedance model we consider a simple cell geometry with two planar electrodes.
This results in exact analytical expressions which elucidate the full parameter dependence of the complex resistance.
We give particular attention to the electrolyte which is described with a thermodynamically consistent theory.
Our theory describes a concentrated and non-ideal binary electrolyte with convection.
We use the Poisson equation which naturally describes charged surface layers.
These layers cause the standard capacitive response in impedance spectroscopy.
The Poisson equation is rarely used in literature \cite{Zhou2005,Moya2016} as most studies simplify the equation system and assume local electroneutrality.
They then model interface capacitances phenomenologically \cite{Legrand2014}.
Our impedance model also takes into account the SEI as a surface film covering the electrode.
We assume \ins{electrolyte} transport in the SEI pores and gain insights into the nature of lithium-ion transport through the SEI by comparing our model with a recent experiment\cite{Wohde2016}.

We briefly summarize our theory based model, and outline our calculation procedure for impedance in \cref{sec:transtheo}.
These calculations are presented in \cref{sec:calc} and discussed in \cref{sec:discussion}.
In \cref{sec:resimp:exp}, we validate our impedance model with a comparison to experimental data. 
Finally, in \cref{sec:summ} we give our conclusion.

\fsection[sec:transtheo]{Theory}
\begin{figure}
\includegraphics[width=\columnwidth]{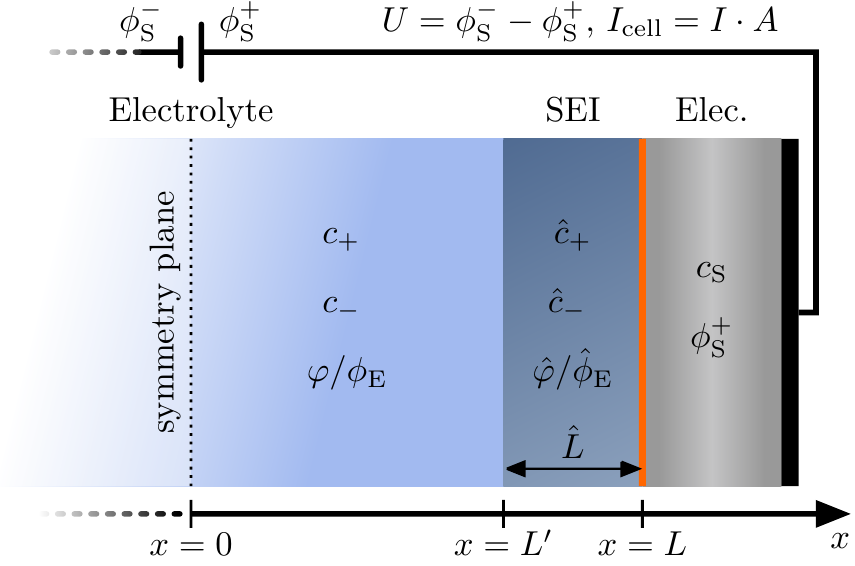}
\caption{\label{fig:cell}
Sketch of the symmetric cell used for the impedance calculation listing relevant variables of each phase.
The SEI thickness $\Lsei$ determines $\Lely=\Ltot-\Lsei$.
The orange boundary between SEI and electrode marks the location of the interface reaction.
}\end{figure}
We calculate the half-cell impedance of the symmetric cell depicted in \cref{fig:cell}.
It consists of two identical planar electrodes which are separated by the SEI and a binary electrolyte. 
This setup represents a common electrochemical cell, e.g., two lithium metal electrodes with LP30 electrolyte.
We describe liquid phases such as the electrolyte and the pore space of the SEI with and without the assumption of local electroneutrality.
Additionally, we consider a simplified model without SEI in each of these scenarios. %\DIFaddbegin \DIFadd{(without what? Please specify, electroneutrality?)}\DIFaddend .
Thus, we discuss a total of four impedance models.
In this way we guide the reader through calculations and discussions as the model complexity increases.

We perform a virtual experiment to calculate the impedance response of our model cells.
To this aim, we apply an oscillating potential or current.
Specifically, we choose a boundary condition for which the temporal progression of the ``applied'' quantity is proportional to $e^{i \omega t}$.
Here, $i$ is the imaginary unit and $\omega=2\pi f$ is a fixed frequency.
We then calculate the corresponding response for this frequency, i.e., current or potential.
All governing equations are linearized for this calculation such that a real-valued solution can be obtained easily from the complex one.
We find the general solution for each primary variable listed in \cref{fig:cell}.
General solutions are a linear superposition of multiple partial solutions because of the linear nature of the problem.
The correct linear combination follows from physical boundary conditions.
Considering these constraints naturally results in a linear system of equations.
Its solution gives the half cell impedance with the conventional definition
\begin{align}
\label{eq:imp}
Z(\omega) = \frac{U}{2I} = \frac{1}{2}\cdot\frac{\PhiS^- - \PhiS^+}{\jint +\jdls},
\end{align}
where the sign in the definition of the voltage difference $U$ considers the difference between technical and physical current.
$\jint$ is the current density corresponding to the rate of the interface reaction.
$j_\indexS$ describes charge that moves between the electrodes to screen charged surface species, see \cref{sec:imp:qdls}.
\Cref{eq:imp} gives the impedance in $\Omega\,$m$^2$ because $\jint$, $j_\indexS$, and $I$ are current densities.
Division of $Z(\omega)$ by $A$, the cross section area of the cell, results in the actual cell resistance.

\fsubsection[sec:trans]{Transport Theory}
We describe transport in the electrolyte phase with a theory derived by Schammer et al. \cite{Schammer2019} based on previous works in refs \citenum{Latz2015,Latz2011,Stamm2017,Clark2017}.
The theory is discussed in the Electronic Supporting Information, see \cref{A-sec:transtheo,A-sec:binary}.
It describes the fluxes of a binary electrolyte consisting of cations, anions and neutral solvent molecules (labelled with \textit{subscript} $\indexP$, $\indexM$, and $\indexN$).
Two independent flux expressions are sufficient to describe the motion of this mixture relative to the center-of-mass velocity $\vv$, e.g.,
\begin{align}
\label{eq:flux1}
\Pflux{\indexA} &= - \sum_{\indexB=\pm} \matD_{\indexA\indexB} \gradv{\cA} - \frac{\tA \cond}{\zA F} \gradv{\PhiE},
\end{align}
where $\indexA=\indexPM$.
This representation is well suited to describe a general electrolyte.
If we assume local electroneutrality, however, we choose the anion flux $\NM$ and the ionic current $\Jcom$ as independent fluxes 
\begin{subequations}
\label{eq:flux2}
\begin{align}
\Pflux{\indexM} &= - \Dsalt \gradv{\cM} + \frac{\tM\Jcom}{\zM F}, \\
\Jcom           &= \frac{\N\tM\cond}{\zP F } \dxdy{\mutsalt}{\csalt}\gradv{\cM} - \cond \gradv{\tvPhi}, \label{eq:flux2b}
\end{align}
\end{subequations}
where $F$ is the Faraday constant and $\N=\frac{\nP+\nM}{\nP\nM}$.
Note that $\Pflux{\indexA}$ and $\Jcom$ are flux and current densities.
Fluxes in \cref{eq:flux1,eq:flux2} are driven by gradients of concentration ${\cA}$,  electric potential $\PhiE$, and effective electrochemical potential $\tvPhi=\PhiE + \frac{\mutP}{\zP F}$.
Effective quantities are marked with a tilde and appear frequently in this work.
They originate from the description relative to the center-of-mass velocity.
The effective chemical potential of cation $\mutP$ and salt $\mutsalt$ are directly related to the conventional quantities, see \cref{A-eq:basif:eff,A-eq:basic:dmusaltdcsalt}.
Transport in the electrolyte is parametrized by the salt diffusion coefficient $\Dsalt$, conductivity $\cond$, and the transference number $\tP$ $(\tP+\tM=1)$.
The diffusion matrix $\bm{\matD}$ with entries $\matD_{\indexA\indexB}$ in \cref{eq:flux1} is determined by these three parameters, see \cref{A-sec:saltdiff}.
Transport parameters are related to the more fundamental Onsager coefficients by the chemical potentials $\muA$.
We use the standard definition of the chemical potentials,
\begin{align}
\label{eq:muA}
\muA = RT \ln\left(\frac{\fA \cA}{\cAEQ}\right), \hspace{.5cm}\indexA=\pm,\indexSalt,
\end{align}
where $\cAEQ$ are the reference concentrations.
The activity coefficients $\fA$ describe the non-ideal behaviour of species $\indexA$ and are related to the thermodynamic coefficient $\thermofac_\indexA = 1 + \pxpy{\ln \muA}{\ln \cA}$, see \cref{A-sec:chempot}.

The center-of-mass velocity $\vv$ is used to express the complete flux expressions
\begin{align}
\label{eq:restflux}
\NA^* = \NA + \cA \vv.
\end{align}
Below, the superscript $*$ labels quantities and parameters that are associated with the complete flux expressions.
These flux expressions are used in mass balance equations
\begin{align}
\label{eq:mbeq}
\partial_t \cA = - \divv \NA^*,
\end{align}
which determine the temporal evolution of a concentration with the corresponding flux density.
As outlined in \cref{A-sec:incomp}, we consider incompressibility as an additional constraint to express $\vv$ \cite{Horstmann2013a,Stamm2017}.

\fsubsection[sec:lin]{Linearization of Model Equations}
Impedance measurements are performed around an equilibrium state, the reference state.
They capture the linear response of this system to an applied potential/current which is oscillating at a given frequency.
Because the reference shall not be perturbed, the applied voltage/current must be small.
In our analytical approach, these oscillations are chosen to have an infinitesimal amplitude.
As a result, any deviation from the reference state in our virtual measurement becomes infinitesimal.
Therefore, all governing equations can be linearized around the reference state.
Considering \cref{eq:flux1,eq:flux2}, the most simple such reference state has a constant concentration and potential distribution
\begin{subequations}
\begin{align}
\cPM(x)		&= \cPMEQ = \nPM \csaltEQ,\\
\PhiE(x)	&= \PhiEEQ,\text{ etc.}
\end{align}
\end{subequations}
Henceforth, we mark all quantities referring to the reference state with the subscript $\indexEQ$.
We refer the interested reader to \cref{A-sec:lin}, where the linearization procedure is described in detail.
Linearization of the flux expressions relative to the center-of-mass velocity results in
\begingroup
\allowdisplaybreaks
\begin{subequations}
\label{eq:fluxlin}
\begin{align}
\label{eq:fluxlin1}
\Pflux{\indexA} 	&= - \sum_{\indexB=\pm} \matD_{\indexA\indexB} \gradv{\deltacB} - \frac{\tA \cond}{\zA F} \gradv{\deltaPhiE}, \\
\label{eq:fluxlin2}
\Pflux{\indexM} 	&= - \Dsalt \gradv{\deltacM} + \frac{\tM\Jcom}{\zM F}, \\
\label{eq:fluxlin3}
\Jcom			&= \frac{\N\tM\cond}{\zP F} \dxdy{\mutsalt}{\csalt} \gradv\deltacM- \cond \gradv{\deltatvPhi}.
\end{align}
\end{subequations}
\endgroup
Linearizing the full flux expressions given by \cref{eq:restflux} results in 
\begingroup
\allowdisplaybreaks
\begin{subequations}
\label{eq:fluxlinr}
\begin{align}
\label{eq:fluxlin1r}
\Pflux{\indexA}^* 	&= - \sum_{\indexB=\pm} \matD_{\indexA\indexB}^* \gradv{\deltacB} - \frac{\tA^* \cond}{\zA F} \gradv{\deltaPhiE} + \cA \vvoff, \\
\label{eq:fluxlin2r}
\Pflux{\indexM}^* 	&= - \Dsalt^* \gradv{\deltacM} + \frac{\tM^*\Jcom}{\zM F}  + \cM \vvoff, \\
\label{eq:fluxlin3r}
\Jcom^*			&= \frac{\N\tM\cond}{\zP F} \dxdy{\mutsalt}{\csalt} \gradv\deltacM- \cond \gradv{\deltatvPhi} = \Jcom,
\end{align}
\end{subequations}
\endgroup
where, $\vvoff$ is a constant offset velocity.
Note that \cref{eq:fluxlin3r} is identical to \cref{eq:fluxlin3} because the charge density in our reference state is zero.
2
The linearized flux expressions have three significant properties.
\textit{Firstly}, all original variables are replaced with the corresponding deviation variables.
They consider the deviation from the reference state
\begin{subequations}
\begin{align}
\cA \rightarrow\,\, &\deltacA = \cA-\cAEQ, \\
\PhiE \rightarrow\,\, &\deltaPhiE = \PhiE-\PhiEEQ,\text{ etc.}
\end{align}
\end{subequations}
\textit{Secondly}, all quantities beside these deviation variables are constant after linearization.
This not only applies to transport parameters but also to the concentrations $\cA$ and partial derivatives such as $\dxdy{\mutsalt}{\csalt}$.
These quantities are consistently evaluated at the reference state.
We therefore omit the corresponding notation in each linearized expression.
\textit{Thirdly}, a new set of apparent transport parameters ($\Dsalt^*$, $\matD_{\indexA\indexB}^*$, and $\tA^*$) consistently replaces the original ones in the linearized full flux expressions.
These quantities combine diffusion/migration and convection in a single diffusion/migration term.
This is a result of linearizing the expression for the center-of-mass velocity.
\Cref{A-eq:labtransbin} relates the apparent transport parameters to the parameters used in flux expression relative to the center-of-mass velocity.

\fsubsection[sec:interface]{Interface Reaction}
We use a linearised Butler-Volmer rate expression to describe the interface reaction rate
\begin{align}
\label{eq:linbv}
\jint = \frac{\PhiI_\indexlin}{\RI}.
\end{align}
Here, $\mathcal{R}$ is the interface resistance parameter of our model.
It is inversely proportional to the exchange current density.
Note that \cref{eq:linbv} does not depend on the charge transfer coefficient and the electrolyte/electrode concentration.
These dependencies are part of the non-linearized rate expression \cite{Rubi2003,Latz2013,Bazant2013a}.
They vanish because the expression is linearized at the reference state where $\PhiI=0$.
The linearized overpotential is equal to
\begin{align}
\label{eq:lineta}
\PhiI_\indexlin 	&= \deltaPhiS - \delta\vPhi^\indexBulk - \pxpy{U}{\cS} \deltacS,
\end{align}
see ref. \citenum{Latz2013}.
This expression takes into account the electrode potential $\deltaPhiS$, the electrochemical potential in the electrolyte $\deltavPhi^\indexBulk$, and the concentration of intercalated particles in the electrode $\deltacS$.
The label ``$\indexBulk$'' indicates that the evaluation of $\deltavPhi$ is non-trivial in the case of spatially resolved double layers.
For simplicity we restrict ourselves to metallic electrodes, i.e., $\pxpy{U}{\cS}=0$, in the main text.
The impact of an intercalation electrode is discussed in \cref{A-sec:calcS}.
In our definition, $\eta$ is negative for intercalation or plating processes.

\fsubsection[sec:sei]{SEI Model}
Experimental and theoretical studies report that SEI is at least partially porous \cite{Harris2013,Michan2015,Lu2014,Single2016,Single2017}.
Our recent findings suggest that solvent molecules are effectively immobilized within these pores \cite{Single2018,Horstmann2018}.
However, this result does not apply to smaller and more mobile lithium ions.
They are also charged and subject to large electric forces.
We follow this idea in this work and model the SEI with nano-sized pores.
These pores are filled with electrolyte and enable charge transport through the surface film.
Parameters, quantities and variables in the SEI pores are marked with a hat.
We use porous electrode theory to describe transport in this pore space \cite{Newman1962,Newman1975,Doyle1993a,Newman2004}.
This means that we employ the same flux expressions that are used for the electrolyte phase, see \cref{eq:fluxlin}.
However, the original bulk transport parameters are replaced with effective ones 
\begin{subequations}
\label{eq:transeff}
\begin{align}
\tDsalt 	&= \frac{\porosity}{\tortuosity} \Dsalt,\\
\tconduc	&= \frac{\porosity}{\tortuosity} \conduc.
\end{align}
\end{subequations}
Parameters $\porosity$ and $\tortuosity$ (porosity and tortuosity) capture the morphology of the SEI.
They are constant in space and time.
Additionally, we introduce $\ttP$, a dedicated cation transference number for the SEI phase.
This is motivated by findings of Popovic et al. \cite{Popovic2016}
They show that the lithium transference number of a liquid electrolyte can be increased and become close to one if the anion species is immobilized in a mesoporous structure.

\fsubsection[sec:boundary]{Boundary Conditions}
The binary electrolyte is in contact with the electrodes which take up lithium ions only.
Therefore, the anion flux density $\NM^*$ vanishes at the electrode interface.
At the same time, $\NP^*$ is equal to the interface reaction rate $\frac{\jint}{\zP F}$.
For the electrodes at $x=\Ltot$, we obtain the following boundary conditions for the fluxes relative to the center-of-mass velocity
\begin{align}
\label{eq:basic:boundrayconvlin}
\NA(\Ltot )=- \frac{\jint}{\zP F} \cdot \left\{
\begin{array}{ll}
1 - \frac{\rhoP}{\rho}   &\indexA=\indexP,       \\
-\frac{\zP}{\zM} \frac{\rhoP}{\rho}       &\indexA=\indexM.
\end{array}
\right. 
\end{align}
Here, $\rhoP$ and $\rho$ are the cation mass density and the mass density of the electrolyte.
\Cref{A-sec:incomp,A-sec:binlin} contain a detailed derivation of the expression above.

\fsection[sec:calc]{Theory of Impedance Spectroscopy}
The most common simplification in the modeling of electrochemical systems is the assumption of local electroneutrality.
We use this assumption for impedance calculations in \cref{sec:imp:neutral} (neutral models).
These calculations are then repeated without the electroneutrality assumption in \cref{sec:imp:nonneutral} (non-neutral models).
We discuss all impedance results in \cref{sec:discussion}.

\fsubsection[sec:imp:neutral]{Electroneutral Impedance}
Local electroneutrality means that the charge density $\varrho$ is zero and that the \ins{ionic} current $\Jcom$ is constant in space.
This assumption also implies that charge does not accumulate at interfaces, therefore the double layer screening charge $Q_\indexS$ and the corresponding current $j_\indexS$ vanish.
In this case, we apply an oscillating cell current $I(t)=I_\indexEQ \eiomt$.
This is convenient because electroneutrality implies $\Jcom=I$ such that \cref{eq:fluxlin3,eq:fluxlin2r} can be used to solve for $\deltacM$ and $\deltavPhi$.

We first calculate the impedance without considering SEI.
Thus, the electrolyte phase spans form $-\Ltot$ to $\Ltot$ and is in direct contact with the electrode.
We add SEI in \cref{sec:neutralsei}.

In the first step we insert the linearized flux expression for $\NM^*$ into the mass balance equation of the anion concentration \cref{eq:mbeq}.
This results in a linear partial differential equation in $\deltacM$,
\begin{align}
\label{eq:imp:neutralode}
\partialt \deltacM = \Dsalt^* \divgradv{\deltacM} ,
\end{align}
as $\divv \Jcom=0$.
We solve it with an exponential Ansatz in $x$ and $t$, i.e. $\deltacM\propto e^{ikx}e^{i\omega t}$.
Only anti-symmetric solutions in $x$ contribute to the impedance calculation, see \cref{eq:imp}.
The solution of \cref{eq:imp:neutralode} then becomes
\begin{align}
\label{eq:imp:neutralsolution}
\deltacM = C \eiomt \sin{\left(k x\right)},
\end{align}
where $C$ is a coefficient and the wave number $k$ is given by the dispersion relation,
\begin{align}
\label{eq:imp:neutralk}
k = \left( 1-i \right) \sqrt{\frac{\omega}{2\Dsalt^*}} .
\end{align}
The inverse of $k$ describes the spatial width of salt concentration oscillations at a given frequency $\omega$.
We determine $C$ with the flux boundary condition for the anion species, see \cref{eq:basic:boundrayconvlin},
\begin{align}
\label{eq:resimp:constneutral}
C\eiomt 	&=\frac{I_\indexEQ \eiomt}{\zM F} \frac{\rho}{\rhohatN}  \frac{1}{\Dsalt^*} \left( \tM - \frac{\rhoP}{\rho} \right) \frac{1}{k \cos(k\Ltot)}.
\end{align}
The extrapolated density $\rhohatN$ is given by  $\rhohatA=\frac{\MA}{\vA}$ ($\MA$ is the molar weight and $\vA$ is the partial molar volume of species $\indexA$).
We find that $C$ is proportional to the amplitude of the applied current $I_0$.

Next, we calculate the deviation of the electrochemical potential $\deltavPhi$ at $x=\Ltot$.
This quantity is needed to express the rate of the interface reaction with \cref{eq:lineta}.
We obtain it by integrating \cref{eq:fluxlin3r}.
However, first, we express the effective electrochemical potential $\tvPhi$ in this equation with $\vPhi$, the conventional one.
\Cref{A-eq:basic:mut,A-eq:chempoteff} relate these quantities,
\begin{align}
\label{eq:tvphi}
\tvPhi &= \vPhi - \frac{\MP}{\MN} \frac{\muN}{\zP F}.
\end{align}
We eliminate the chemical potential of the solvent $\muN$ with the Gibbs-Duhem relation, see \cref{A-eq:basic:gibbsduhem2}.
The differential version of \cref{eq:tvphi} then becomes
\begin{align}
\pxpy{\deltatvPhi}{x} = \pxpy{\deltavPhi}{x} + \frac{\N}{\zP F} \frac{\rhoP}{\rho}  \dxdy{\mutsalt}{\csalt} \pxpy{\deltacM}{x}, \label{eq:dvarphin}
\end{align}
where, electrolyte density $\rho$ and cation density $\rhoP$ are constant and evaluated at the reference state.
We use \cref{eq:dvarphin} in \cref{eq:fluxlin3r} and rearrange for $\gradv \vPhi$.
Integration from $0$ to $x$ results in 
\begin{align}
\label{eq:varphin}
\deltavPhi	&= -\frac{I x}{\cond} + \frac{\N}{\zP F} \left(\tM - \frac{\rhoP}{\rho} \right)\dxdy{\mutsalt}{\csalt} \cdot \deltacM(x).
\end{align}
The anti-symmetry of concentration and electrochemical potential implies that both $\deltacM$ and $\deltavPhi$ vanish at $x=0$.

\fsubsubsection[sec:reactpot]{Interface Reaction}
We describe the interface reaction rate $\jint$ with the linearized Butler-Volmer expression given by \cref{eq:linbv}.
In the electroneutral model, we evaluate the electrochemical potential at the interface $\deltavPhi^\indexBulk=\deltavPhi(\Ltot)$.
The current between the electrodes $I$ and the interface reaction rate $\jint$ are related by
\begin{align}
I=-\jint = -\PhiI_\indexlin \mathcal{R}^{-1},
\end{align}
where the sign considers the orientation of the interface.
Inserting the linearized overpotential $\eta_\indexlin$ from \cref{eq:lineta} gives the potential of the electrode
\begin{align}
\label{eq:phisn}
\deltaPhiS^+ = \deltaPhiS(+\Ltot) =- I \mathcal{R}  + \deltavPhi(\Ltot).
\end{align}

\fsubsubsection[sec:neutralimp2]{Impedance}
Considering the symmetry of the solution implies $U=\PhiE^--\PhiE^+=-2\deltaPhiE^+$.
\Cref{eq:imp} then implies $Z=\deltaPhiS^+/I$ if $j_\indexS=0$ is considered.
Therefore, we obtain the complex impedance by inserting \cref{eq:varphin} in \cref{eq:phisn}, considering \cref{eq:imp:neutralsolution,eq:resimp:constneutral}, and dividing by $I$.
We find that $Z$ is the sum of three distinct contributions
\begin{align}
\label{eq:imp:zneutral}
Z(\omega) = \RE + \Rint + \underbrace{\RD\cdot\frac{\tan\left(k\Ltot\right)}{k\Ltot}}_{\ZD}.
\end{align}
The complex impedance depends on frequency $\omega$ through the dispersion relation $k(\omega)$, see \cref{eq:imp:neutralk}.
Here, the ohmic contributions $\RE$, $\Rint$ and $\RD$ are constant and do not depend on frequency,
\begin{subequations}
\label{eq:imp:rneutral}
\begin{align}
\RE 	&= \frac{\Ltot}{\conduc}, \label{eq:re}\\
\Rint 	&= \mathcal{R}, \label{eq:rint}\\
\RD 	&= \frac{-\N}{\zP \zM F^2} \frac{\Ltot}{\Dsalt^*} \left(\tM - \frac{\rhoP}{\rho}\right)^2 \underbrace{\frac{\rho^2}{\rhoN\rhohatN}}_{\mathcal{M}} \dxdy{\musalt}{\csalt} \label{eq:rd}.
\end{align}
\end{subequations}
We attribute $\RE$ and $\Rint$ to the resistance of the electrolyte and the interface reaction.
$\RD$ and $\ZD$ describe a finite-length Warburg impedance or Warburg Short ($W_\mathrm{S}$) \cite{warburg1901ueber,Neumann1899}.
This is the impedance increase of the electrolyte as salt concentration gradients form at low frequencies.
The cation density $\rhoP=\MP\cP$ and solvent density \ins{$\rhoN$}$=\MN\cN$ as well as $\rhohatN=\MN\vN\inv$ determine $\RD$ in \cref{eq:rd}.
The relative cation density ${\rhoP}/{\rho}$ appears as a correction of the transference number $\tM$. It vanishes in the dilute limit.
We rewrite the factor $\mathcal{M}$ as a function of $\alpha = \csalt \vSalt$ and $\beta=\rhohatSalt/\rhohatN$ ($\rhohatSalt = M_\indexSalt/\vSalt$)
\begin{align}
\label{eq:imp:convectionfactor}
\mathcal{M}=\frac{\rho^2}{\rhoN \rhohatN}
= (1-\alpha) \left(1 + \frac{\alpha\beta}{1-\alpha}\right)^2.
\end{align}
This factor has a non-linear dependence on the salt concentration through $\alpha$.
It approaches one in the dilute limit and diverges at the maximum salt concentration $\vSalt\inv$.%, see \cref{fig:imp_convectionfactor}.

\fsubsubsection[sec:neutralsei]{Solid-Electrolyte Interphase}
The SEI covers negative electrodes in Li-ion batteries.
In this subsection, we take SEI into account as a porous surface film, see \cref{sec:sei} and \cref{fig:cell}.
As described in \cref{sec:transtheo}, we use porous electrode theory to describe transport in this interphase \cite{Newman1962,Newman1975,Doyle1993a,Newman2004}.
To this aim, we replace the transport parameters in \cref{eq:fluxlin3r,eq:fluxlin2r} with effective parameters for the SEI phase.
We use these expressions in the modified mass balance equation, \cref{eq:mbeq}, and consider SEI porosity
\begin{align}
\partial_t (\porosity \deltatcM) =  \tDsalt^* \divgradv{(\porosity \deltatcM)}.
\end{align}
This equation describes the temporal evolution of the anion concentration in the SEI pores.
In analogy to \cref{eq:imp:neutralsolution}, an exponential Ansatz results in the dispersion relation 
\begin{align}
\label{eq:resimp:ksein}
\ksei 	= \left(1-i\right) \sqrt{\frac{\tortuosity \omega}{2 \Dsalt^*}} = \left(1-i\right) \sqrt{\frac{\porosity \omega}{2 \hat{\Dsymbol}_\indexSalt^*}}.
\end{align}
Here, the symmetry does not simplify the solution.
Thus, the anion concentration in the SEI pores contains leftmoving and rightmoving waves
\begin{align}
\label{eq:deltacMgen}
\deltatcM = \eiomt\left( \hat{C}^+ e^{i \ksei x} + \hat{C}^- e^{-i \ksei x} \right).
\end{align}
The concentration in the electrolyte phase is given by \cref{eq:imp:neutralsolution}, also in this case.
We now determine the three coefficients $C$, $\hat{C}^+$, and $\hat{C}^-$ with interface boundary conditions.
Electrolyte and SEI phase share the interface at $x=\pm\Lely$.
Both phases must have the same salt concentration at this point, i.e., $\deltacM(\Lely) = \deltatcM(\Lely)$.
The anion flux must also be continuous, i.e., $\NM^*(\Lely) = \hat{\Nsymbol^*}_\indexM(\Lely)$.
As the convective flux is identical in both phases we can use $\NM(\Lely)=\hat{\Nsymbol}_\indexM(\Lely)$ instead.
The anion flux must satisfy the boundary condition at the electrode interface, see \cref{eq:basic:boundrayconvlin}.
We combine these three constraints in a linear system of equations
\begin{align}
\label{eq:resimp:neutralsei}
&\left(\begin{array}{ccc}
\sin\left(k\tL\right)	&e^{i \ksei \tL}				&e^{-i \ksei \tL} 				\\
k \cos\left(k\tL\right)	&-i\ksei \frac{\porosity}{\tortuosity} e^{i \ksei \tL}	&i \ksei  \frac{\porosity}{\tortuosity} e^{-i \ksei \tL}	\\
0				&i\ksei  \frac{\porosity}{\tortuosity} e^{i \ksei \Ltot}	&-i \ksei  \frac{\porosity}{\tortuosity} e^{-i \ksei \Ltot}
\end{array} \right)
\vec{C}
\notag \\&\hspace{2cm}= 
\frac{I\emiomt}{\zM F \Dsalt}
\left(\begin{array}{c}
0\\
\tM -\ttM\\
\ttM - \frac{\rhoP}{\rho}
\end{array}\right),
\end{align}
where $\vec{C}=(C,\hat{C}^+,\hat{C}^-)\transpose$ is the coefficient vector.
These equations are solved analytically, see \cref{A-eq:sol:neutralsei}.%, see the attached file ``impedance\_SEI\_neutral.nb''.

Next, we calculate $\vPhi(\Ltot)$, the electrochemical potential at the electrode interface.
In analogy to \cref{eq:varphin}, we rearrange \cref{eq:fluxlin3} and find $\deltavPhi$ by integration and by considering \cref{eq:dvarphin}
\begin{align}
\label{eq:resimp:zneutralsei}
&\deltavPhi(\Ltot) = 	-\frac{I}{\conduc}\left(\tL + \frac{\tortuosity}{\porosity} \Lsei\right) 
				+ \frac{\N}{\zP  F}\dxdy{\mutsalt}{\csalt} \cdot \notag \\
&\hspace{.6cm}\left[ \left(\tM - \ttM \right) \deltatcM(\tL) 
 + \left(\ttM - \frac{\rhoP}{\rho} \right) \deltatcM(\Ltot) \right].
\end{align}
We integrate over the electrolyte and the porous SEI phase which have different transference numbers $\tM$ and $\ttM$.
The concentration deviations $\deltatcM$ at $x=\tL$ and $x=\Ltot$ are given by \cref{eq:deltacMgen} and the solution of \cref{eq:resimp:neutralsei}.
Next, we insert $\deltavPhi(\Ltot)$ in \cref{eq:phisn} to express the electrode potential $\deltaPhiS^+$.
As in \cref{sec:neutralimp2}, the half-cell impedance $Z$ is given by $\deltaPhiS^+/I$,
\begin{align}
\label{eq:resimp:ZneutralSEI}
Z(\omega) = & \REp + \RSEI + \Rint + \ZDsei + \ZDp,
\end{align}
where $\tRE = \Lely/\conduc$ is the adjusted resistance of the electrolyte and $\RSEI = \Lsei/\tconduc$ is the resistance of the SEI.
Here, quantities labeled with $'$ replace the corresponding quantities in the model without SEI \cref{eq:imp:zneutral}.
The interface resistance $\Rint$ is still given by \cref{eq:rint}.
$\ZDsei$ and $\tZD$ are stated in \cref{eq:zdsei}.
They describe the impedance increase due to the build-up of salt concentration profiles in SEI and electrolyte phase.
The length and diffusion coefficient of each phase ($\Lely$, $\Lsei$ and $\Dsalt^*$, $\tDsalt^*$) determine a characteristic frequency for this process in each domain.
\begin{subequations}
\label{eq:zdsei}
\begin{align}
\ZD' 	= & \frac{\Lely\Theta}{\Dsalt^*}\bigg(2\left(\ttM-\tM\right)\left(\ttM - \frac{\rhoP}{\rho}\right) \sec\left(\ksei\Lsei\right) \notag \\
&+  \left(\ttM-\tM\right)^2 + \left(\ttM-\frac{\rhoP}{\rho}\right)^2 \bigg) 
\cdot \frac{\tan(\,k \Lely)}{\Psi\cdot k\Lely} \label{eq:resimp:zwbbulk2}\\
\ZDsei 	= &\frac{\Lsei\Theta}{\tDsalt^*}\left(\ttM - \frac{\rhoP}{\rho} \right)^2 \cdot \frac{\tan(\ksei\Lsei)}{\Psi \cdot \ksei\Lsei}, \label{eq:resimp:zwbsei2}
\end{align}
\end{subequations}
where 
\begin{subequations}
\begin{align}
\label{eq:imp:theta}
\Theta &= \frac{-\N}{\zP\zM F^2 } \frac{\rho^2}{\rhoN \rhohatN}  \dxdy{\musalt}{\csalt}, \\
\Psi &= 1 - \porosity \sqrt{\tortuosity\inv} \, \tan{(k\Lely)} \,\tan{(\ksei\Lsei)}.
\label{eq:imp:psi}
\end{align}
\end{subequations}

\fsubsection[sec:imp:nonneutral]{General Impedance}
In this section, we calculate the impedance without the assumption of local electroneutrality.
Without electroneutrality, the number of independent concentrations in the electrolyte increases by one ($\cP$ and $\cM$).
We use the Poisson equation to account for this new variable
\begin{align}
-\frac{\varrho}{\epszr} = \divgradv \PhiE = \divgradv \deltaPhiE.
\end{align}
It relates the electrostatic potential in the electrolyte $\PhiE$ with the ionic charge density $\varrho$.
The direct appearance of $\PhiE$ in one of the primary equations makes it reasonable to use $\PhiE$ as a variable instead of the electrochemical potential $\vPhi$.
As a consequence, we use a different set of flux expressions for the non-neutral system (\cref{eq:fluxlin1r} instead of \cref{eq:fluxlin2r,eq:fluxlin3r}).
Inserting \cref{eq:fluxlin1r} into a mass-balance equation for $\cP$ and $\cM$ results in
\begin{align}
\label{eq:imp:main_odea}
\partialt \deltacA = \sum_{\beta=\indexPM} \DAB \divgradv \deltacB + \frac{\tA^*\cond}{\zA F} \divgradv \deltaPhiE .
\end{align}
because $\vvoff$ is constant.
Now, we use the Poisson equation to eliminate the electric field
\begin{align}
\label{eq:resimp:insertpoisson}
\partialt \deltacA = \sum_{\beta=\indexPM} \left( \DAB \divgradv \deltacB - \frac{\tA^*\cond}{\epszr}\frac{\zB}{\zA} \deltacB \right) .
\end{align}
Using the vector $\vecdeltac=(\deltacP,\deltacM)\transpose$ and the matrix
\begin{align}
\matT^* 
&= \frac{\conduc}{\epszr} \left( \begin{array}{cc}
\tP^* 			& -\nP\nM\inv \tP^*\\
-\nM \nP\inv\tM^* 	&\tM^*
\end{array}\right),
\end{align}
we write \cref{eq:resimp:insertpoisson} in matrix form
\begin{align}
\label{eq:imp:main_ode}
\partialt \vecdeltac = \bm{\matD}^* \divgradv \vecdeltac - \matT^* \vecdeltac,
\end{align}
where differential operators are applied element wise.
\Cref{eq:imp:main_ode} is a coupled linear ODE in $\vecdeltac$.
We solve this equation with an exponential ansatz
\begin{align}
\vecdeltac = \veceta \cdot \eiomt \eikx .
\end{align}
$\veceta$ is a coefficient vector and $\omega$ is a fixed frequency.
This results in an algebraic matrix equation
\begin{align}
0 = \bigg(k^2\matI + \underbrace{{\bm{\matD^*}}^{-1}\left( \matT^* + \iom\matI\right)}_{\matA} \bigg) \veceta,
\end{align}
where $\matI$ is the identity matrix.
If $\veceta$ is an eigenvector of $\matA$ with eigentwert $\lambda$, we obtain
\begin{align}
\label{eq:klambda}
k^2 = -\lambda.
\end{align}
The matrix $\matA$ has two eigenvectors $\vecetaP$ and $\vecetaM$ with eigenvalues $\lambda_\indexO$ and $\lambda_\indexT$.
Note that these eigenvalues and eigenvectors depend on the frequency $\omega$.
Below, we scale the eigenvectors of $\matA$ so that their second entry equals 1, i.e., $\vecetaA = \left( \eta_\indexA , 1 \right)\transpose$.
We then obtain four possible solutions for $k$
\begin{align}
k_\indexA^\pm = \pm \sqrt{-\lambda_\indexA}, \hspace{.5cm} \indexA=\indexO,\indexT.
\end{align}
Due to the superposition principle, we obtain the general solution for the concentration deviation
\begin{align}
\vecdeltac 	%&= \eiomt \sum_{\indexA=\indexO,\indexT} \vecetaA \cdot \left( \CAP \epikAx + \CAM \emikAx \right) \notag \\
				&= \eiomt \sum_{\indexA=\indexO,\indexT} \vecetaA \GammaA \label{eq:cgeneral} .
\end{align}
The solution is determined by four coefficients $\CApm$ which are contained in the function $\GammaA(x)$% and $\GammaAp(x)$
\begin{subequations} 
\label{eq:imp:Gamma}
\begin{align} 
\GammaA(x) &= \GammaA = \CAP \epikAx + \CAM \emikAx, \\
\GammaAp(x) &= \GammaAp = \CAP \epikAx - \CAM \emikAx,
\end{align}
\end{subequations} 
where $\indexA=\indexO,\indexT$.
These functions are introduced for readability.

The electrostatic potential $\PhiE$ in the electrolyte is differentially directly linked to the free charge density via the Poisson equation.
We use the concentrations $\deltacP$ and $\deltacM$ given by \cref{eq:cgeneral} to express $\varrho = F\sum_\indexA \zA\deltacA$.
We then insert it in the Poisson equation and obtain
\begin{align}
\label{eq:phigeneral}
\deltaPhiE = \eiomt F \left( \sum_{\alpha=\indexO,\indexT} \PiA \GammaA + \Pp x + \Pz \right),
\end{align}
by integrating twice.
Here, $\Pp$ and $\Pz$ are integration constants and $\PiA$ is
\begin{align}
\label{eq:imp:PiA}
\PiA = \frac{\zP\etaA + \zM}{\epszr\kAsq},\hspace{.5cm}\indexA=\indexO,\indexT.
\end{align}

Six coefficients $\CPpm$, $\CMpm$, $\Pp$, and $\Pz$ define the general solution of $\vecdeltac$ and $\deltaPhiE$ for a given frequency $\omega$.
We determine these constants with physical boundary conditions in \cref{sec:solnosei,sec:imp:nonneutralsei}.
To this aim, the expressions for $\vecdeltac$ and $\deltaPhiE$ from \cref{eq:cgeneral,eq:phigeneral} are inserted into the flux expression \cref{eq:fluxlin1}.
We then obtain the linearized flux expression relative to the center-of-mass velocity
\begin{align}
\label{eq:fluxgensol}
\NA = -\eiomt \left( \sum_{\indexB=\indexO,\indexT} \Omega_{\indexA\indexB} \GammaBp + \frac{\tA\cond}{\zA} \Pp \right),
\end{align}
where the 2x2 matrix $\bm{\Omega}$ with indices $\alpha=\indexPM$ and $\beta=\indexO,\indexT$ is given by
\begin{align}
\label{eq:basic:matom}
\Omega_{\indexA\indexB} &= i \kB \left( \frac{\cond\tA}{\zA} \PiB + \DAPcom \etaB + \DAMcom \right).
\end{align}

\begin{figure}
\includegraphics[width=\columnwidth]{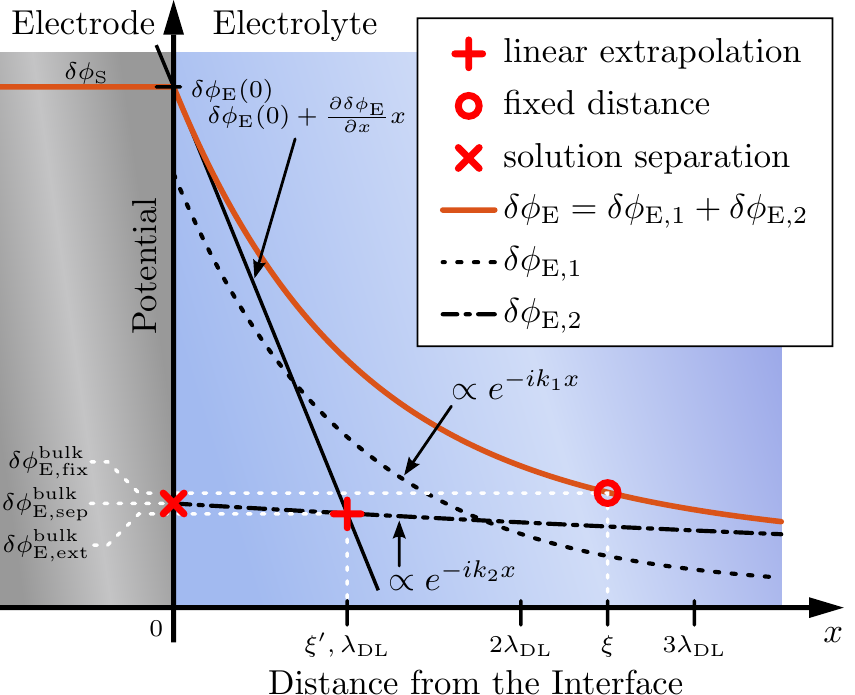}
\caption{\label{fig:interface}
Illustration of the potential deviation close to the interface with three methods to determine $\deltaPhiE^\indexBulk$.
The red line shows the spatial dependence of $\deltaPhiE$ which can be separated into two contributions $\deltaPhiE = \delta\phi_{\indexE,\indexO} + \delta\phi_{\indexE,\indexT}$.
The spatial dependence of these parts is given by $e^{-i \kP x}$ and $e^{- i \kM x}$ respectively, see \cref{eq:cgeneral,eq:phigeneral}.
For illustrative purposes, $\kM$ has been chosen equal to $\kP/10$. 
}\end{figure}

\fsubsubsection[sec:imp:qdls]{Double Layer Screening Current}
Usually, solid electrodes are electronically highly conductive.
We assume that this conductivity is infinite which is a good approximation for metal electrodes or graphite.
Thus, the potential within the electrode is spatially constant, see \cref{fig:interface}.
Therefore, the electric potential has a kink at the interface between the electrode and the electrolyte.
This implies charge accumulation at the interface according to Gauss' law.
This charge is provided by free charge carriers (electrons) from the electrode.
%It screens external electronic forces from the double layer in the electrolyte and is referred to as the double layer screening charge.
It is determined by the potential gradient in the electrolyte at the interface
\begin{align}
Q_\indexS = \epszr \pxpy{ \PhiE}{x}  \bigg|_{x=\Ltot}= \epszr \pxpy{\deltaPhiEom}{x}  \bigg|_{x=\Ltot}.
\end{align}
The current which supplies these charges is obtained from the temporal derivative of $Q_\indexS$
\begin{align}
\jdls &= \pxpy{Q_\indexS}{t} = i\omega\epszr \pxpy{\deltaPhiEom}{x} \bigg|_{x=\Ltot} \notag  %\notag\\&= i \omega \epszr \eiomt F \left(\sum_\indexA i\kA \PiA \GammaAp(\pm L) + \Pp \right) 
\\&=  i\omega\epszr \eiomt F \left( \sum_{\alpha=\indexO,\indexT} i\kA \PiA \GammaAp(\Ltot) + \Pp \right) \label{eq:imp:jscreen}.
\end{align}

\fsubsubsection[sec:resimp:dispersion]{Dispersion Relation}
\begin{figure}[t]
\includegraphics[width=\columnwidth]{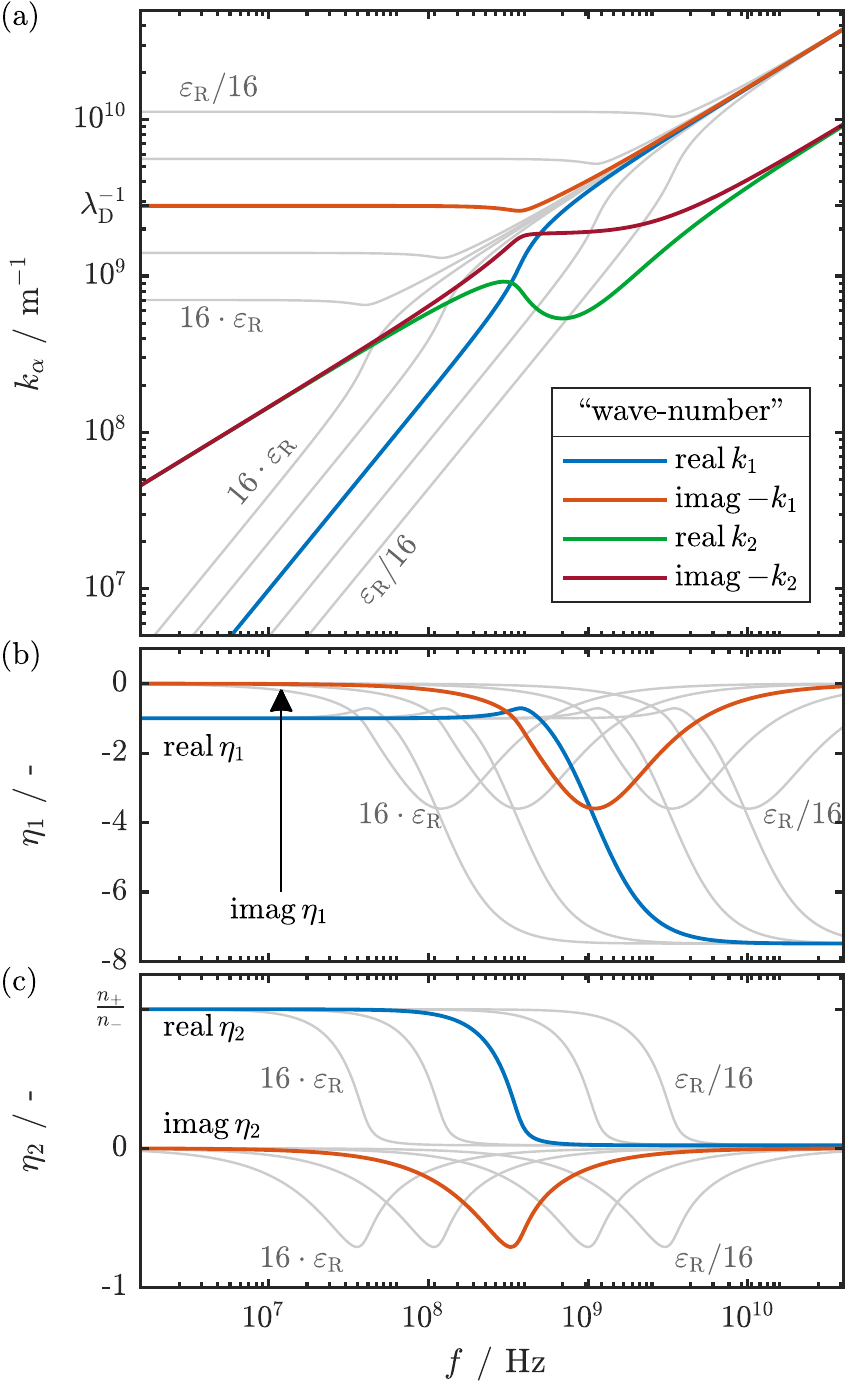}
\caption{\label{fig:kvsfreq}
Frequency dependence of $\kA$ and $\etaA$ for a monovalent salt (LiPF$_6$ EC/DMC).
The parameters used are listed in \cref{A-tab:param}.
The grey lines illustrate the dependence on $\epsr$ which we change in multiples of $4$ around the base value, i.e., $\epsr=1.96$, $7.85$, $31.41$, $125.6$, and $502.6$.
}\end{figure}
The relation between the wave-numbers $\kA$ and the frequency $\omega$ is called dispersion relation.
We find this expression with the eigenvalues $\lambda_\indexA$ of the matrix $\matA$, see \cref{eq:klambda}.
The analytic solution is presented in \cref{A-sec:dispersion}.
We illustrate the dispersion relation and both eigenvectors \ins{for LP30 electrolyte} in \cref{fig:kvsfreq}.
Here, a distinct physical meaning emerges for each wave number/eigenvector pair at frequencies below a transition frequency.
We identify this frequency as $f_\mathrm{trans}=\frac{\conduc}{2\pi \epszr}$ which lies between $10^6-10^8\,$Hz for reasonable parameters.
For all frequencies $f$ below this value we find that $\kM$ aligns with $k$, the wave number of the neutral solutions given by \cref{eq:imp:neutralk}.
Simultaneously, $\etaM$ attains the constant value $\nP/\nM=-\zM/\zP$.
Therefore, $\vecetaM \GammaM$ describes charge-neutral salt concentration oscillations on the system scale.
This eigenvalue/vector-pair is referred to as the ``far-field'' eigenvalue/vector-pair.

In contrast, imag$\,\kP$ quickly attains a constant value below the critical frequency $f_\mathrm{trans}$.
We refer to the inverse of this value ($|\kP|\inv$) as the double layer length $\lambdaDL$.
This quantity is typically equal to a few \AA{} in standard lithium-ion batteries.
We obtain this value by evaluating $\kP$ at $\omega=0$ (see \cref{A-sec:lambdadl}),
\begin{align}
\label{eq:lambdaDL}
\lambdaDL = \sqrt{-\frac{\epszr RT}{\zP\zM F^2} 	\frac{\N\thermoSalt}{\csalt}	{\thermosplit \left(1-\thermosplit\right)} }.
\end{align}
The value of $\etaP$ at frequencies below $f_\mathrm{trans}$ corresponds to a non-electroneutral electrolyte.
Therefore, $\vecetaP \GammaP$ describes diffuse and charged double layers which decay exponentially with $\lambdaDL$.
Solutions of this eigenvalue/vector-pair become only relevant near the interfaces and are referred to as ``near-field'' solutions below.

Electrochemical impedance measurements are usually not performed with frequencies larger than $10^6\,$Hz.
This means that all experimentally relevant frequencies are smaller than $f_\mathrm{trans}$.
The relative permittivity does not depend on the frequency in this frequency range as well.
For instance, in the case of ethylene carbonate, $\epsr$ begins to change at frequencies larger than $10^9\,$Hz \cite{Payne1972}.
%In the derivation above $\epsr$ is assumed to be constant, therefore all solutions are only valid in the corresponding frequency range.
We therefore conclude that $\etaP$, $\etaM$, and $\kP=-i\lambdaDL\inv$ are constant in the relevant frequency range.
Additionally, $\kM$ can be approximated by $k$, see \cref{eq:imp:neutralk}.

\fsubsubsection[sec:resimp:modelcorr]{Real Double Layer and Interface Capacity}
An important distinction that sets our model apart from other similar models is the non-ideal electrolyte that we consider.
Non-ideal behaviour is captured by the thermodynamic coefficient $\thermoSalt$ and the asymmetry factor $\thermosplit$, see \cref{A-eq:basic:tau}.
Additionally, we consider ionic interactions and ionic association with the Onsager matrix, specifically the transference numbers.
However, our theory does not consider the finite size of individual ions and molecules.
%As a result, $\lambdaDL$ depends only on bulk properties of the electrolyte.
%This results in a 
This can result in a wrong prediction of the double layer thickness $\lambdaDL$.
We take such errors into account by manually adjusting $\lambdaDL$ with the dimensionless parameter $\zeta$
\begin{align}
\label{eq:resimp:lambdaDLcorr}
&\lambdaDL \rightarrow \zeta \cdot  \lambdaDL,&
&\text{and}&
\tlambdaDL \rightarrow \zeta \cdot  \tlambdaDL.&
\end{align}
This modification adjusts the double layer capacity $\Cint$ which determines $\fint$, the resonance frequency of the interface reaction, see \cref{eq:resimp:fi}.

In realistic systems, $\Cint$ includes capacitive contributions from a layer of specifically adsorbed ions.
However, such a layer is not considered in our model, see \cref{eq:imp:phis}.
\Cref{eq:resimp:lambdaDLcorr} also corrects for this simplification.

Our impedance model assumes a reference state without a diffuse layer at the interface.
This is similar to the assumption that the electrodes are polarized to the potential of zero charge and requires that the electrodes are polarized to a specific potential.
If the electrodes are polarized to any other potential, charged double layers will be part of the reference state.
These double layers become several nm thick in ionic liquids \cite{Hoffmann2018}.
In this scenario, our theory does not predict the double-layer thickness correctly.

In conclusion, some model simplifications result in the incorrect prediction of $\fint$, the resonance frequency of the interface reaction.
We use \cref{eq:resimp:lambdaDLcorr} in \cref{sec:resimp:exp} for comparing our model with experimental impedance data.

\fsubsubsection[sec:resimp:interface]{Interface Reaction}
The interface reaction is driven by the linearized overpotential given in \cref{eq:lineta}.
This expression depends on the electrochemical potential in the electrolyte
\begin{align}
\label{eq:varphilin}
\deltavPhi^\indexBulk = \deltaPhiE^\indexBulk + \frac{1}{\zP F}\pxpy{\muP}{\cP} \deltacP^\indexBulk.
\end{align}
In electroneutral models, we evaluate these quantities directly at the interface as these theories do not resolve charged double layers.
%The issue of the spatial extend of the interface reaction does not naturally appear in neutral models and is often neglected.
%However, in non-neutral theories, the interface and the bulk electrolyte are separated by a charged double layer which is few \AA{} thick.
As illustrated in \cref{fig:interface}, charged double layers can contribute significantly to the concentration and potential deviation despite being only a few \AA{} thick \cite{PhysRevE.83.061507}.
%Classical continuum models cannot resolve the exact reaction position.
The combination of a Butler-Volmer rate expression and a locally electroneutral electrolyte is a well established method.
Therefore, agreement with the neutral theory is a prerequisite for the non-neutral model.
This can only be achieved if the double layer contributions to concentration and potential deviation in \cref{eq:varphilin} are not included in the definition of the ``bulk'' values.
\Cref{fig:interface} illustrates three methods which achieve this:
 \begin{enumerate} 
\item\fsubsubsubsection{Fixed distance}
We can evaluate the deviation variables at a fixed distance $\xi$ in front of the interface %to obtain the bulk values
\begin{align}
\delta\phi_{\indexE,\mathrm{fix}}^{\indexBulk} = \deltaPhiE(\xi).
\end{align}
This has two disadvantages.
First, an additional parameter $\xi$ is introduced by this definition.
Note that the bulk values are used to define the overpotential which is in turn used to express the reaction rate.
This reaction rate is used in the flux boundary condition at the interface.
Then, the boundary conditions depend on variables which are not evaluated at the boundary itself.
\item\fsubsubsubsection{Linear extrapolation}
This problem can be bypassed by using linear extrapolation to obtain the bulk value
\begin{align}
\delta\phi_{\indexE,\mathrm{ext}}^{\indexBulk} = \deltaPhiE(0) + \xi' \cdot \pxpy{\deltaPhiE}{x}\bigg|_{x=0}.
\end{align}
In this method, both the deviation variable and its derivative are evaluated at the interface.
Note that this definition also requires one additional parameter $\xi'$.
\item\fsubsubsubsection{Solution separation}
This method can be used if the deviation variable can be uniquely decomposed into a part that describes the diffuse layer in front of the interface and a bulk contribution.
Then, the near-field contribution can be subtracted from the value at the interface to obtain the far-field value
\begin{align}
\delta\phi_{\indexE,\mathrm{sep}}^{\indexBulk} = \deltaPhiE(0) - \delta\phi_{\indexE,\indexO}(0) = \delta\phi_{\indexE,\indexT}(0).
\end{align}
%Here, the bulk value equal to the ``rest'' evaluated at the interface.
The advantage of this definition over the other methods is that no additional interface length $\xi$ has to be defined.
It also clearly connects to neutral models.
 \end{enumerate} 
\Cref{fig:interface} illustrates that the bulk values $\delta\phi_{\indexE,\mathrm{fix}}^{\indexBulk}$, $\delta\phi_{\indexE,\mathrm{ext}}^{\indexBulk}$, and $\delta\phi_{\indexE,\mathrm{sep}}^{\indexBulk}$ nearly coincide for a reasonable choice of the additional parameters. Here, we assume $\xi\gg\lambdaDL$ for the first method and $\xi'=\lambdaDL$ for the second method.

As discussed in \cref{sec:resimp:dispersion}, our solution is clearly divided into a near-field and a far-field part in the relevant frequency range.
We therefore use the solution separation method which yields simple expressions for the boundary values, simplifying the analytical calculations,
\begin{subequations}
\label{eq:imp:bulk}
\begin{align}
\vecdeltacom^{\text{ }\indexBulk} 	%&= \vecdeltacom(\Ltot)- \eiomt \vecetaP \GammaP(\Ltot) \notag \\
					&= \eiomt \vecetaM \GammaM(\Ltot),\\
\deltaPhiEom^\indexBulk 		%&= \deltaPhiEom(\Ltot) - \eiomt F \PiP \GammaP(\Ltot) \notag \\
					&= \eiomt F \left( \PiM \GammaM(\Ltot) + \Pp \Ltot + \Pz \right). 
\end{align}
\end{subequations}

In the neutral system we connect electrode and electrolyte potential with the rate expression, see \cref{eq:phisn}.
This is possible because the reaction rate $\jint$ is equal to $I$, the external current between the electrodes. 
However, in the non-neutral system charge can accumulate in the diffuse layer and in the double layer screening charge $\Qdls$, see \cref{sec:imp:qdls}.
%Then, $\jint$ and $I$ are no longer the same and a new equation is needed to relate electrode and electrolyte potential.
Then, $I$ is equal to $\jint+\jdls$ and a new equation is needed to relate electrode and electrolyte potential.
We assume that the potential \ins{deviation} is a continuous function of space so that
\begin{align}
\label{eq:imp:phis}
\deltaPhiS^+ =  \deltaPhiE(\Ltot).
\end{align}
This is illustrated in \cref{fig:interface}.
Most theories of the electrochemical double layer consider a diffuse layer and a layer of specifically adsorbed ions \cite{Newman2004}.
Specifically adsorbed ions have at least partially lost their solvation shell and are in direct contact with the interface.
This layer can have a net charge and a dipole moment.
By using \cref{eq:imp:phis} we neglect the dynamics of these quantities.

Modeling the dynamics of the inner Helmholtz plane  is beyond the scope of this work.
An interested reader is referred to dedicated works on this subject \cite{Luck2016}.

We now express the linearized overpotential $\eta_\indexlin$, see \cref{eq:lineta}, with \cref{eq:varphilin,eq:imp:phis,eq:imp:bulk}
\begin{align}
\label{eq:imp:jintred}
\PhiI_\indexlin &= \eiomt \left( F\PiP\GammaP(\Ltot) - \frac{\etaM}{\zP F} \pxpy{\muP}{\cP} \GammaM(\Ltot) \right).
\end{align}
Here, we consider the simple concentration dependence of the chemical potentials given by \cref{eq:muA}.
We introduce the dimensionless parameter $\thermosplit$ to connect $\thermoP$ with $\thermoSalt$ in \cref{A-sec:chempot}

\fsubsubsection[sec:solnosei]{Solution without Solid Electrolyte Interphase}
In this subsection, we discuss the cell without SEI.
This means that the electrolyte spans from $-\Ltot$ to $\Ltot$ and is in direct contact with the electrode.
We use the symmetry argument to eliminate three of the six coefficients which define the general solution
\begin{subequations}
\begin{align}
\CP &= \CP^+ = -\CP^-, \\
\CM &= \CM^+ = -\CM^-, \\
\Pz &= 0.
\end{align}
\end{subequations}
As a result, the functions $\GammaA$ and $\GammaAp$ become common trigonometric expressions
\begin{subequations}
\begin{align}
\label{eq:imp:gammasimple}
\GammaA &= 2i \CA \sin\left(\kA x\right), \\
\GammaAp &= 2 \CA \cos \left( \kA x \right) .
\end{align}
\end{subequations}
We insert these functions into the flux expression \cref{eq:fluxgensol} and the equation for the interface reaction rate \cref{eq:linbv,eq:imp:jintred}.
Next, both of these quantities are used to write the flux boundary conditions given by \cref{eq:basic:boundrayconvlin}.
This results in two homogeneous linear equations in the remaining coefficients $\CP,\CM,\PP$.
We write them in matrix form
\begin{align}
\label{eq:imp:matixeq}
\boldcal{S} \vec{C} = 0,
\end{align}
where $\vec{C} = \left(\CP,\CM,\PP\right)\transpose$ is the coefficient vector and $\boldcal{S}$ is a 2x3 matrix given by \cref{eq:imp:solsym}.
\Cref{eq:imp:matixeq} determines $\vec{C}$ with respect to its amplitude.
We give the analytic solution in the \ESI, see \cref{A-eq:sol:nnneutral}. %This solution is used to express the impedance with \cref{eq:imp}.
This solution defines all quantities in \cref{eq:imp}, i.e., $\PhiS^+$, $\jint$, and $j_\indexS$ with \cref{eq:imp:phis,eq:phigeneral,eq:linbv,eq:imp:jintred}.
We therefore use \cref{eq:imp} to calculate $Z$ analytically, see \cref{A-eq:z:nnneutral}.

\fwidebegin
\begin{align}
\label{eq:imp:solsym}
\boldcal{S}
&=-\left(\begin{array}{ccc}
\Omega_{\indexP\indexO} \cdot 2\cos\left(\kP \Ltot\right) &\Omega_{\indexP\indexT}  \cdot 2\cos\left(\kM \Ltot\right)&\frac{\tP\cond}{\zP} \\
\Omega_{\indexM\indexO} \cdot 2\cos\left(\kP \Ltot\right) &\Omega_{\indexM\indexT}  \cdot 2\cos\left(\kM \Ltot\right)&\frac{\tM\cond}{\zM}
\end{array} \right) +\notag \\
&\hspace{1cm}
\frac{1}{\RI}
\left(\begin{array}{ccc}
\frac{1}{\zP}\left(1-\frac{\rhoP}{\rho}\right)	&0			&0\\
			0			&\frac{1}{\zM}\frac{\rhoP}{\rho}	&0
\end{array} \right)
\left(\begin{array}{ccc}
\PiP \cdot 2i \sin\left(\kP \Ltot\right)  & -\frac{\etaM}{\zP F^2} \pxpy{\muP}{\cP} \cdot 2i \sin\left(\kM \Ltot\right)& 0 \\
\PiP \cdot 2i \sin\left(\kP \Ltot\right)  & -\frac{\etaM}{\zP F^2} \pxpy{\muP}{\cP} \cdot 2i \sin\left(\kM \Ltot\right)& 0 \\
0&0&0
\end{array} \right) .
\end{align}
\fwideend

\fsubsubsection[sec:imp:nonneutralsei]{Solution with Solid Electrolyte Interphase}
Next, we transfer the solution of the non-neutral electrolyte to the porous SEI which spans from $x=\Lely$ to $x=\Ltot$.
To account for the morphology of the SEI, we use the effective transport parameters introduced in \cref{sec:sei}.
However, the porosity of the SEI phase also appears in a few specific steps during the calculation.
It enters the Poisson equation,
\begin{align}
\label{eq:poissonmod}
-\frac{\porosity\varrho}{\tepszr} = \divgradv \deltaPhiE.
\end{align}
Here, we replace the charge density in the pores with the averaged charge density in the SEI.
Additionally, we introduce $\tepsr$, the mean permittivity of the SEI.
The porosity also appears in the mass balance equation.
Considering these changes, the modified version of \cref{eq:imp:main_ode} becomes
\begin{align}
\label{eq:imp:sei_ode}
\partialt \vecdeltatc = \porosity^{-1}\hat{\bm{\matD}}^* \divgradv \vecdeltatc - \hat{\matT}^* \vecdeltatc.
\end{align}
We calculate $\hat{\bm{\matD}}^*$ and $\hat{\matT}^*$ in the same way as in the electrolyte phase but use the effective transport parameters for the SEI phase.
However, $\porosity\inv\hat{\bm\matD}^*$ instead of $\bm\matD^*$ is used for the calculation of $\tkA$ and $\tetaA$.
The modified Poisson equation \cref{eq:poissonmod} also affects the definition of $\tPiA$,
\begin{align}
\tPiA = \frac{\porosity \left(\zP \tetaA + \zM\right)}{\epszr\tkA^2}.
\end{align}
The SEI specific set of transport parameters, eigenvalues, and eigenvectors ($\tkA$, $\tetaA$) is used in the definition of $\tGammaA(x)$.
We then find the frequency dependent solutions of concentration and potential deviation
\begin{subequations}
\label{eq:resimp:seisol}
\begin{align}
\label{eq:resimp:seisolc}
\vecdeltacom &= \eiomt \cdot \left\{ \begin{array}{ll}
\sum_\alpha \vecetaA \GammaA	&x\leq \Lely \\
\sum_\alpha \vectetaA \tGammaA 	&x\geq \Lely
\end{array} \right. , \\
\label{eq:resimp:seisolphi}
\deltaPhiEom &= \eiomt F \cdot \left\{ \begin{array}{ll}
\sum_\alpha \PiA  \GammaA	+ \Pp\cdot x			&x\leq \Lely \\
\sum_\alpha \tPiA \tGammaA 	+ \tPp\cdot x 	+ \tPz		&x\geq \Lely
\end{array} \right.
\end{align}
\end{subequations}
for $x$ between $0$ and $\Ltot$.
Compared to the system without SEI, six additional coefficients need to be determined.
Consequently, we consider six additional boundary conditions.
 \begin{enumerate} 
\item Both $\vecdeltacom$ and $\deltaPhiEom$ are continuous $x=\Lely$.
\item The particle fluxes $\NP$ and $\NM$ are continuous at $x=\Lely$.
\item No charge is stored at the interface between the electrolyte and the SEI phase and Gauss's law implies $\epsr \partialx \deltaPhiEom(\Lely) = \tepsr \partialx \delta\tPhiE(\Lely)$ 
 \end{enumerate} 
These six additional equations are linear in the coefficients.
We write them in matrix form with the expanded coefficient vector $\vecC = (\CP$, $\CM$, $\Pp,$ $\tCP^+$, $\tCP^-$, $\tCM^+$, $\tCM^-$, $\tPp$, $\tPz)\transpose$.
This results in the 8x9 matrix $\hat{\boldcal{S}}$ if the two flux boundary conditions at $x=\Ltot$ are considered, see \cref{eq:basic:boundrayconvlin}.
We denote this matrix in the \ESI{} in \cref{A-sec:nnsei}.
The coefficient vector must satisfy
\begin{align}
\label{eq:syssei}
\hat{\boldcal{S}}\vecC = 0.
\end{align}
Because of the increased size of the system of equations we do not perform this calculation analytically.
Instead, we solve \cref{eq:syssei} numerically to obtain the coefficient vector $\vecC$ for each frequency of interest.
We then use this result in the expressions for $\deltaPhiE$, $\jint$, and $j_\indexS$, to calculate the impedance $Z$ with \cref{eq:imp}.

\fsection[sec:discussion]{Discussion}
In this section we analyze and discuss the impedance models derived in the previous section.
We first discuss the models without SEI and compare the differences between the neutral and non-neutral approach in \cref{sec:discimpnosei}.
The impedance models with SEI are then discussed in \cref{sec:resimp:resultSEI}.
An essential part of our analysis is the approximation and simplification of the non-neutral models.
Our simplified models bring out the parameter dependence of the impedance signal over a large parameter range. %We also gain an understanding under what circumstances the common simplifications hold.
The corresponding equivalent circuits are discussed in \cref{sec:resimp:equiv}.

\begin{figure}[t]
\includegraphics[width=\columnwidth]{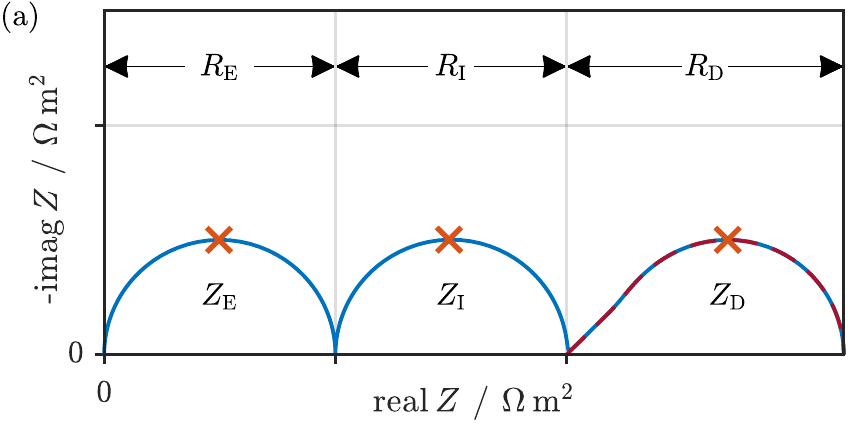}
\includegraphics[width=\columnwidth]{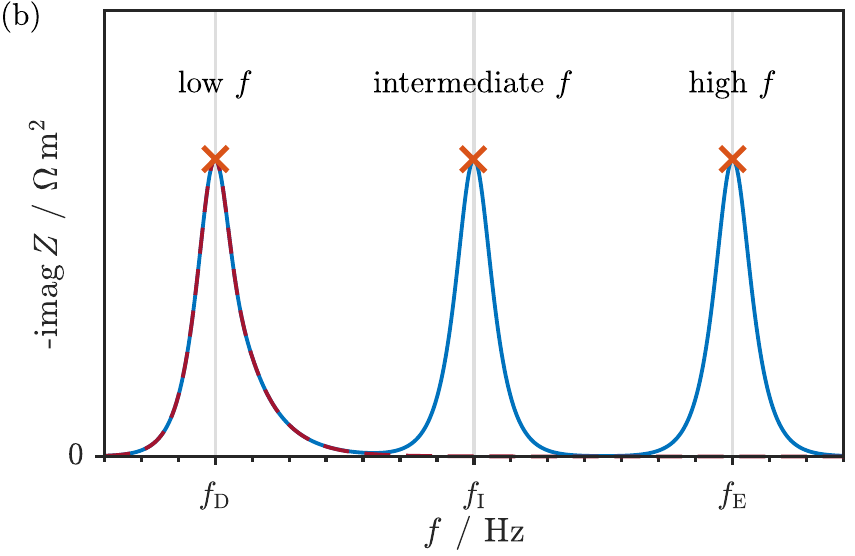}
\caption{\label{fig:impnosei}
Schematic impedance spectrum of the symmetric cell without SEI.
(a) Nyquist plot. (b) Bode plot.
The solid blue line shows the impedance of the non-neutral model whereas the dashed red ones show the impedance of the neutral one.
Crosses mark the resonance frequencies.
}\end{figure}

\fsubsection[sec:discimpnosei]{Impedance - Without SEI}
\Cref{fig:impnosei} shows the impedance of the half-cell without SEI.
It consists of three distinct features, for both, the neutral and the non-neutral model, namely
 \begin{itemize} 
\item the resistance of the electrolyte $\RE$/$\ZE$,% see \cref{eq:re},
\item the interface resistance $\Rint$/$\Zint$,% see \cref{eq:rint},
\item the polarization impedance $\RD$/$\ZD$.%, see \cref{eq:imp:zneutral,eq:rd}.
 \end{itemize} 
We label real impedance contributions that do not depend on frequency with $R$.
In contrast, complex impedance contributions that depend on frequency are labelled with $Z$.
Our neutral model does not predict the frequency dependence of electrolyte and interface resistance.
However, neutral models give the polarization impedance $\ZD$ in \cref{sec:impdiffusion}.
The frequency dependent impedance of interface reaction $\Zint$ and electrolyte $\ZE$ are discussed with our non-neutral model in \cref{sec:resimp:result_simple}.

\fsubsubsection[sec:impdiffusion]{Diffusion Impedance}

\begin{figure}[ht!]
\subfloat[\label{fig:zda}]{}
\subfloat[\label{fig:zdb}]{}
\subfloat[\label{fig:rhofac}]{}
\makebox[\columnwidth][l]{\raisebox{-1cm}{\includegraphics[width=\columnwidth]{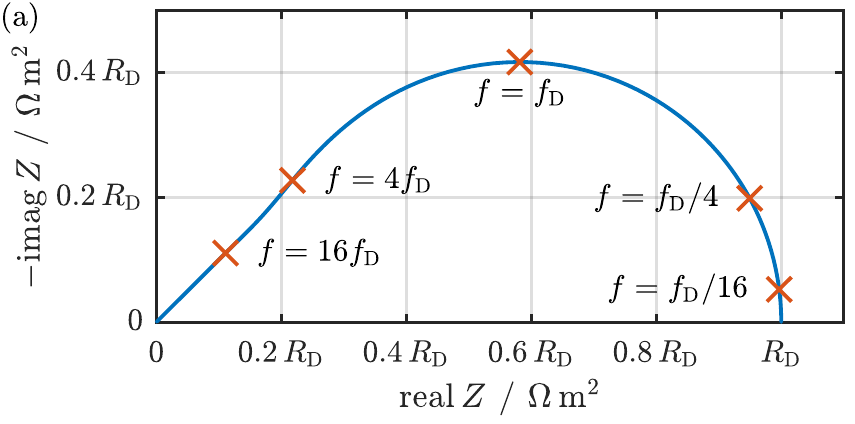}}}
\makebox[\columnwidth][l]{\raisebox{-0cm}{\includegraphics[width=\columnwidth]{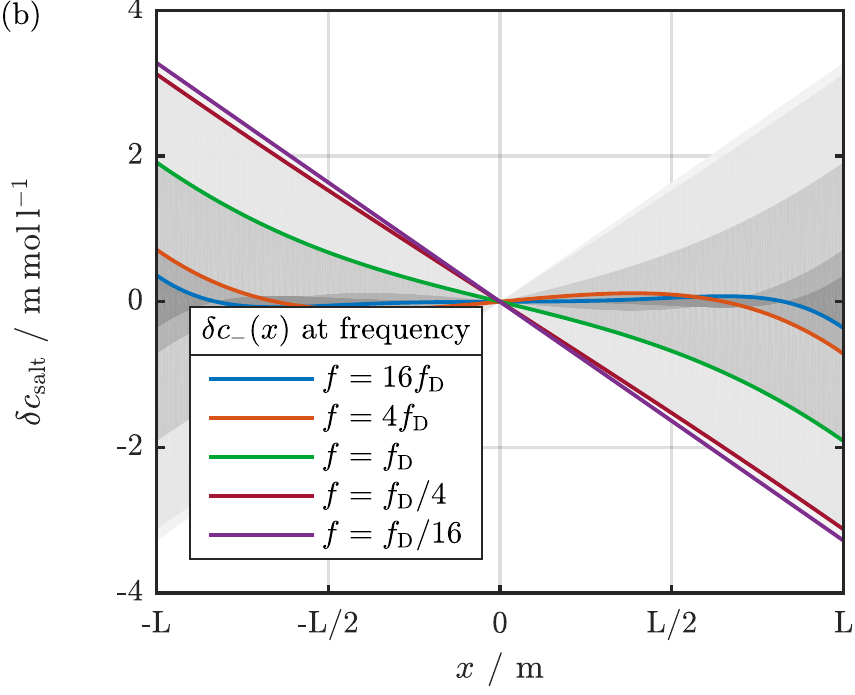}}}
\makebox[\columnwidth][l]{\raisebox{-0cm}{\includegraphics[width=\columnwidth]{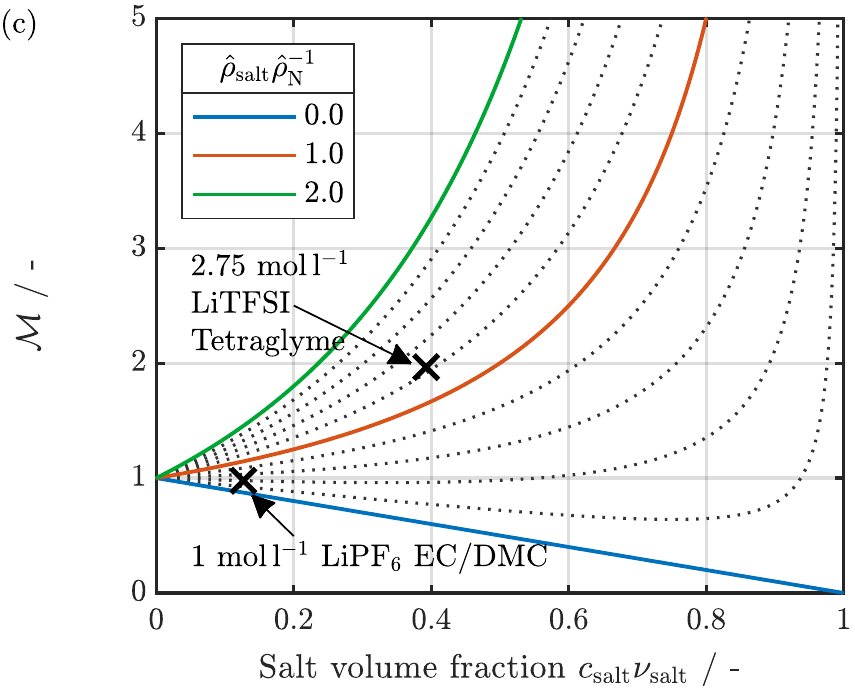}}}
\caption{\label{fig:zd}
(a) Illustration of the finite-length Warburg impedance.
(b) Envelops of the salt concentration profile in the cell for five different frequencies marked in part (a).
The profiles are obtained for the LP30 electrolyte with a current density of $I_0=10\,\mu$A$\,$cm$^{-2}$.
Parameters are listed in \cref{A-tab:param}. 
	(c) The factor $\mathcal{M}$\remove{$=\rho^2 \rhoN\inv \rhohatN\inv$} is a function of $\csalt\vSalt$, see \cref{eq:imp:convectionfactor}. 
$\mathcal{M}$ scales $\RD$, the amplitude of the diffusion impedance, see \cref{eq:rd}.
Parameters to calculate the marked values are listed in \cref{A-tab:param}.
}\end{figure}

The neutral impedance without SEI is given by \cref{eq:imp:zneutral,eq:imp:rneutral}.
The assumption of local electroneutrality is incompatible with charge accumulation at the interface.
Therefore, the only complex and frequency dependent impedance contribution in the neutral model is the diffusion resistance
\begin{align}
\label{eq:zdfin}
\ZD = \RD\, \frac{\tan{\left(k\Ltot\right)}}{k\Ltot}.
\end{align}
This function is illustrated in \cref{fig:zda}, $\RD$ is given by \cref{eq:rd}.
$\ZD$ describes the impedance increase of the electrolyte as salt concentration gradients emerge at low frequencies.
We illustrate these salt concentration gradients for different frequencies in \cref{fig:zdb}.
The label $\ZD$ denotes that this process is governed by salt diffusion.
In literature, it is referred to as a diffusion, Warburg Short ($\mathrm{W}_\mathrm{S}$), and finite-length Warburg impedance.
$\RD$, the amplitude of this effect is determined by numerous parameters, namely the distance between the electrodes \ins{$\Ltot$}, the transference number $\tM$, and the salt diffusion coefficient $\Dsalt^*$.
Note that $\tM$ is referenced to the center-of-mass velocity, whereas $\Dsalt^*$ is an apparent parameter as defined in \cref{sec:lin}.
$\RD$ is proportional to $\dxdy{\musalt}{\csalt}$ which depends on the salt concentration $\csalt$ and the thermodynamic factor $\thermoSalt$.
The relative cation density $\rhoP/\rho$ appears as correction of the transference number $\tM$ in concentrated electrolytes.
Additionally, $\RD$ is proportional to the factor \ins{$\mathcal{M}=$}$\rho^2 \rhoN^{-1}\rhohatN^{-1}$ which is rewritten in \cref{eq:imp:convectionfactor}.
\ins{$\mathcal{M}$ is equal to 0.977 and 1.965 for the LP30 and LiTFSI electrolyte, see \cref{A-tab:param}.}
We illustrate this factor in \cref{fig:rhofac}, showing that it diverges if the salt concentration reaches its theoretical maximum.
It approaches unity for dilute solutions.
In conclusion, the amplitude of the diffusion resistance has a complex parameter dependence.

In contrast, we find that the frequency dependence of $\ZD$ is simply given by the frequency dependence of $k$, see \cref{eq:imp:neutralk}.
The wave number $k$ depends only on $\Dsalt^*$, the apparent salt diffusion coefficient.
The characteristic timescales of the diffusion impedance $\ZD$ also depend on $\Ltot$, the distance between the electrodes.
We calculate the resonance frequency of $\ZD$ numerically and obtain the following approximation
\begin{align}
\label{eq:resimp:fcritdiff}
f_\mathrm{D} \approx \frac{1.2703 \Dsalt^*}{\pi \Ltot^2}.
\end{align}
\Cref{fig:zdb} illustrates the oscillations in the concentration profiles at various frequencies close to $f_\mathrm{D}$.

$\ZD$ is typically not observed in modern batteries or test cells.
It is covered by other contributions such as diffusive processes in intercalation electrodes.
Therefore, $\ZD$ is best observed if non-intercalation electrodes are used, i.e.\ins{,} metallic lithium.
Another challenge in measuring $\ZD$ are the low frequencies that have to be considered (sub mHz).
Such measurements take a long time and require great care to avoid the initial state from being perturbed.
$f_\mathrm{D}$ can be shifted towards higher values by reducing the distance between the electrodes, see \cref{eq:resimp:fcritdiff}.
However, this reduces the amplitude of the effect.

\fsubsubsection[sec:resimp:result_simple]{Electrolyte and Interface Impedance}
Our neutral models predict a real valued and frequency independent resistance for electrolyte and interface ($\RE$ and $\Rint$), see \cref{eq:re,eq:rint}.
We therefore use the non-neutral impedance model without SEI, see \cref{sec:solnosei}, to discuss $\ZE$ and $\Zint$.
The full expression for $Z$ is given by \cref{A-eq:z:nnneutral}.
In contrast to the electroneutral case, both, the resistance of the electrolyte $\ZE$ and the interface resistance $\Zint$, are frequency dependent.
Their Nyquist plots have the common semicircle shape, see \cref{fig:impnosei}.

The full expression for $Z$ is too intricate for a direct analysis.
However, we find a simplified approximation when three conditions are met.
Firstly, the distance between the electrodes is larger than the double layer thickness, i.e., $L\gg\lambdaDL$.
Secondly, the interface reaction is parametrized such that its resonance frequency $f_\mathrm{I}$ is smaller than $f_\mathrm{trans}=\frac{\conduc}{2\pi\epszr}$, see \cref{sec:resimp:dispersion}.
We show below  that these assumptions are equivalent to $\RI\gg\lambdaDL/\conduc$.
This allows us to approximate $\kP$, $\etaP$, and $\etaM$ as constants as discussed in \cref{sec:resimp:dispersion}.
Finally, we assume $\fD\ll\fE$ which allows us to \ins{approximate \cref{A-eq:z:nnneutral} with} three distinct contributions,
\begin{align}
\label{eq:znoseiimpfull}
Z(\omega) = \ZE + \Zint + \ZD.
\end{align}
%Here, $\ZD$ is the diffusion resistance derived in the neutral model, see \cref{sec:impdiffusion}.
\ins{The expression we obtain for $\ZD$ aligns with the diffusion resistance derived with} the neutral model in \cref{sec:impdiffusion}.
\ins{$\ZE$ and $\Zint$ are given by
\begin{subequations}
\begin{align}
\label{eq:resimp:zhf} \ZE &= \frac{\Ltot}{\conduc + i \omega \epszr}, \\
\label{eq:resimp:zi} \Zint &= \frac{\lambdaDL}{\lambdaDL/\RI+ i \omega \epszr}.
\end{align}
\end{subequations}
}
\ins{Alternatively, we can derive these expressions with equivalent circuits.}
%We now derive the frequency dependent expression for electrolyte and interface impedance $\ZE$ and $\Zint$ with equivalent circuits.
Both electrodes form a parallel-plate capacitor filled with electrolyte, a polarizable medium. %Its capacitance per square meter equals $\frac{\epszr}{2L}$.
In parallel to this capacitance, the electrolyte acts as an ohmic resistor.
The capacity and ohmic resistance of these elements are equal to $\CE = {\epszr}/{\Ltot}$ and $\RE={\Ltot}/{\conduc}$.
We then obtain \cref{eq:resimp:zhf} according to Kirchhoff's rule.
The resonance frequency of this semicircle is given by
\begin{align}
\label{eq:resimp:fcritE}
f_\mathrm{E} = \frac{\conduc}{2\pi\epszr} = f_\mathrm{trans}.
\end{align}
Only the conductivity $\conduc$ and the dielectric constant $\epsr$ of the electrolyte influence this frequency.
Note that $f_\mathrm{E}$ is equal to $f_\mathrm{trans}$, see \cref{sec:resimp:dispersion}.
It marks the transition from the simple low frequency behavior of $\kA$ and $\etaA$ to the more complicated high frequency one.
It is in the $100-1000\,$MHz range for common parameters, making it too large to be observed in an electrochemical impedance measurement.
Consequently, the electrolyte impedance $\ZE$ is typically treated as a constant and purely ohmic contribution.

The interface resistance $\Zint$ corresponds to the charge-transfer reaction.
We obtain $\Zint$ with a parallel circuit of the interface capacitance $\Cint=\epszr / \lambdaDL$ (see \cref{A-eq:resimp:cdl}) and the interface resistance $\Rint=\RI$
The resonance frequency of the interface reaction is equal to
\begin{align}
\label{eq:resimp:fi}
\fint = \frac{\lambdaDL}{2\pi\epszr \RI}
\end{align}
and depends on $\lambdaD$, $\epsr$, and $\RI$.
The complex parameter dependence of the double layer thickness $\lambdaDL$ is given in \cref{eq:lambdaDL}.
We discuss eventual shortcomings and corrections of this prediction in \cref{sec:resimp:modelcorr}.

The agreement between the sum of the three simplified expressions and the full expression for $Z$ is excellent.
It is retained even if the resonance frequencies of the interface reaction and the diffusion impedance overlap.
Differences between our simplification and the full expression become relevant if the electrodes are less than $10 \lambdaDL\sim$5$\,$nm apart.
Furthermore, we observe deviations if $\fint$ or $\fD$ are larger than $\fE/10$.
Such conditions do not appear in standard battery cells.
To conclude, the impedance without SEI is given by two conventional semicircles and the Warburg diffusion element which is derived with the neutral model.

\fsubsection[sec:resimp:resultSEI]{Impedance - With SEI}
\begin{figure}
\includegraphics[width=\columnwidth]{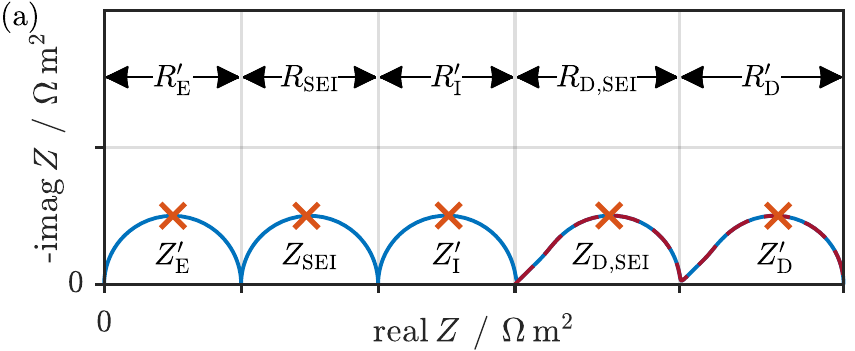}
\includegraphics[width=\columnwidth]{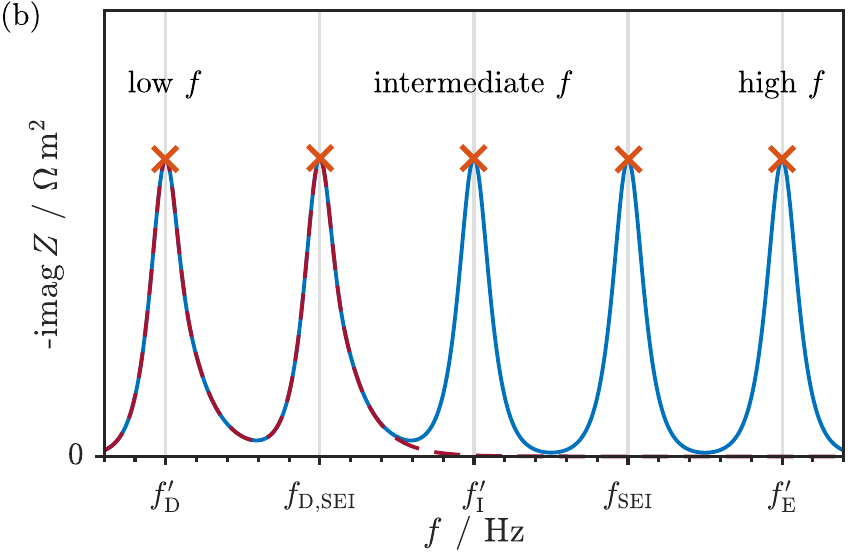}
\caption{\label{fig:impsei}
Schematic impedance spectrum of the symmetric cell with SEI.
(a) Nyquist plot. (b) Bode plot.
The solid blue line shows the impedance of the non-neutral model whereas the dashed red ones show the impedance of the neutral one.
Crosses mark the resonance frequencies.
}\end{figure}
The presence of the  SEI complicates the impedance calculation so that an analytical solution of the non-neutral impedance is no longer feasible.
Therefore, this model is only solved numerically.
However, we find that it is well approximated by the sum of five distinct impedance contributions,
\begin{align}
\label{eq:zimpfull}
Z(\omega) =  \ZEp + \Zsei + \Zintp + \ZDsei + \ZDp.
\end{align}
We illustrate this result in \cref{fig:impsei}.
It shows that the SEI adds two impedance features to our model. %Two impedance features emerge if the SEI is taken into account.
One semicircle represents the ionic transport resistance of the surface film which we label $\Zsei$.
Furthermore, a second diffusion resistance, $\ZDsei$, appears.
This is a consequence of modeling charge transport through the SEI with a liquid electrolyte.
Liquid electrolytes allow the formation of concentration gradients as opposed to solid electrolytes.
All three remaining impedance contributions in \cref{eq:zimpfull} are marked with a $'$.
This indicates that the expressions given in \cref{sec:discimpnosei} must be adjusted to account for the slightly modified geometry ($\Ltot\rightarrow\Lely$).

\fsubsubsection[sec:resimp:result_elyteinterface]{Electrolyte and Interface Impedance} 
We revisit the impedance contributions in \cref{eq:zimpfull} that appear in the impedance model without SEI, see \cref{sec:discimpnosei}.
The presence of SEI reduces the size of the electrolyte phase.
It is now given by $\Lely=\Ltot-\Lsei$ where $\Lsei$ is the thickness of the SEI.
We consider this and modify the expression for the electrolyte resistance
\begin{align}
\REp = \frac{\Lely}{\conduc},& &\text{and}& &\ZEp = \frac{\Lely}{\conduc + i\omega \epszr}.
\end{align}
This change does not affect the corresponding resonance frequency, $\fEp=\fE$.

We denote the double-layer thickness in the SEI pores with $\tlambdaDL$.
It is different from the double-layer thickness in the bulk electrolyte $\lambdaDL$.
Thus, the expression for the interface impedance becomes
\begin{subequations}
\begin{align}
\label{eq:resimp:zisei}
\Zintp &= \frac{\tlambdaDL}{\tlambdaDL/\RI+ i \omega \epszr}, \\
\label{eq:resimp:fisei}
\fintp &= \frac{\tlambdaDL}{2\pi\epszr \RI}.
\end{align}
\end{subequations}
The presence of SEI also affects the diffusion impedance of the electrolyte $\ZD$.
It is replaced by $\ZDp$ which is given by \cref{eq:resimp:zwbbulk2}.
We discuss this in \cref{sec:resimp:result_SEIneutral}.

\fsubsubsection[sec:resimp:result_SEI]{SEI Impedance} 
$\Zsei$ describes the ionic impedance of the SEI.
In analogy to \cref{sec:resimp:result_simple}, we find a good approximation for this expression with an equivalent circuit.
To this aim we model the SEI as a parallel circuit consisting of a capacitor and an ohmic resistance,
\begin{align}
\label{eq:crsei}
\hspace{.5cm}
\Csei &= \frac{\tepszr}{\Lsei}, &\text{and} 	&	&\Rsei &= \frac{\Lsei}{\tconduc} .
\end{align}
The capacitive and ohmic impedance contribution depends on the SEI thickness $\Lsei$, the conductivity $\tconduc$, and the relative permittivity $\tepsr$.
This corresponds to the common assumption that SEI resistance depends mostly on its thickness \cite{Peled1979,Broussely2001}.
We then obtain
\begin{align}
\label{eq:resimp:zhf2}
\hspace{.5cm}
\Zsei = \frac{\Lsei}{\tconduc + i \omega \epszr} = \frac{\Lsei}{\frac{\porosity}{\tortuosity}\conduc + i \omega \epszr}
\end{align}
from Kirchhoff's law.
The resonance frequency of this semicircle is given by
\begin{align}
\label{eq:resimp:fsei}
\hspace{.5cm}
\fsei  = \frac{\porosity\conduc}{2\pi\tortuosity\tepszr} = \frac{\porosity}{\tortuosity} \frac{\epsr}{\tepsr} f_\mathrm{E}.
\end{align}

\fsubsubsection[sec:resimp:result_SEIneutral]{Diffusion Impedance in SEI and Electrolyte} 
In this section, we discuss and simplify the expressions for the diffusion resistance of the SEI and the electrolyte phase ($\Zsei$ and $\ZDp$, see \cref{eq:zdsei}).
We calculate these expressions with the neutral impedance model in \cref{sec:neutralsei}.
As discussed in \cref{sec:sei}, we assume that the SEI is nano porous.
This implies that the SEI porosity $\porosity$ is small and suggests that the SEI tortuosity $\tortuosity$ is large.
Therefore, we assume $\frac{\porosity}{\tortuosity}<\frac{\porosity}{\sqrt{\tortuosity}}\ll1$ in the simplifications below.
%Taking into account that $|\tan\left(k\Lely\right)|<1.2$ and $|\tan(\ksei\Lsei)|<1.2$ then implies $\Psi\approx1$, see \cref{eq:imp:psi}.
\ins{Taking into account that both $|\tan\left(k\Lely\right)|$ and $|\tan(\ksei\Lsei)|$ are bounded by $1.2$ then implies $\Psi\approx1$, see \cref{eq:imp:psi}.}
This results in an approximate expression for $\ZDsei$,
\begin{align}
\ZDsei  \approx \frac{\Lsei\Theta}{\tDsalt^*}\left(\ttM - \frac{\rhoP}{\rho} \right)^2 \frac{\tan(\ksei\Lsei)}{\ksei\Lsei}. \label{eq:resimp:zwbsei3}
\end{align}
This equation has the same structure as the expression for $\ZD$ derived in \cref{sec:impdiffusion}.
We therefore transfer the results from \cref{sec:impdiffusion} and apply them to \cref{eq:resimp:zwbsei3}.
In this way, we obtain an approximation for the resonance frequency of the SEI diffusion impedance
\begin{align}
\label{eq:resimp:fcritsei}
f_\mathrm{D,SEI} \approx \frac{1.2703\tDsalt^*}{\pi\Lsei^2} .
\end{align}

We also use $\Psi\approx1$ to simplify the electrolyte diffusion impedance $\ZDp$ which is given by \cref{eq:resimp:zwbbulk2}.
To further simplify this expression, we assume that the resonance frequency of the diffusion resistance in the SEI $\fDsei$ is larger than the resonance frequency of the diffusion impedance in the electrolyte $\fD$.
This implies that $\sec(\ksei\Lsei)\approx 1$ in the relevant frequency range in \cref{eq:resimp:zwbbulk2}, resulting in
\begin{align}
\ZDp \approx \frac{\Lely \Theta}{\Dsalt^*} \left(\tM - \frac{\rhoP}{\rho}\right)^2	 \frac{\tan\,(k \Lely)}{k\Lely}. \label{eq:resimp:zwbbulk2simple2}
\end{align}
This is the same expression as the original expression for $\ZD$ in \cref{eq:zdfin}, besides the replacement of $\Ltot$ with $\Lely$.
%These expressions differ only in the parameter $\Lely$ which consistently replaces $\Ltot$.
We use this similarity to find the equation for the resonance frequency of $\ZDp$,
\begin{align}
\label{eq:resimp:fcritelyp2}
\fDp \approx \frac{1.2703\Dsalt^*}{\pi\Lely^2} .
\end{align}
\Cref{eq:resimp:zwbbulk2simple2,eq:resimp:fcritelyp2} are important results.
They show that the SEI does not influence electrolyte impedance contributions ($\ZEp\approx\ZE$ and $\ZDp\approx\ZD$), if we consider that the SEI is thin ($\Lely\approx\Ltot$).
%This is true unless salt concentration gradients in the SEI and the electrolyte form at similar timescales.

Next, we compare the amplitude of $\ZDsei$ with the amplitude of $\ZDp$.
No approximation is used for this comparison.
We divide \cref{eq:resimp:zwbsei2} by \cref{eq:resimp:zwbbulk2} in the stationary limit $\omega\rightarrow0$
\begin{align}
\label{eq:resimp:compare1}
\frac{\RDsei}{\RDp}  = \frac{\Rsei}{\RE} \left(\ttM - \frac{\rhoP}{\rho}\right)^2 \left(\tM - \frac{\rhoP}{\rho} \right)^{-2} .
\end{align}
Here, we consider that $\frac{ \tortuosity}{\porosity}\frac{\Lsei}{\Lely}$ is equal to $\frac{\Rsei}{\RE}$.
In most impedance experiments the resistance of the electrolyte $\RE$ is found to be smaller than $\Rsei$, the resistance of the SEI \cite{Wohde2016,Steinhauer2017a}.
If we assume $\Rsei>\RE$, we obtain the following inequality
\begin{align}
\label{eq:resimp:compare}
\frac{\RDsei}{\RDp}  > \left(\ttM - \frac{\rhoP}{\rho}\right)^2 \left(\tM - \frac{\rhoP}{\rho} \right)^{-2} .
\end{align}
Thus, if $\ttM\gtrsim\tM$ the diffusion impedance of the SEI would be larger than the diffusion impedance of the electrolyte.
Because this is not observed in experiments, we conclude that the transference number in the SEI pores is different from the bulk value.
Specifically, $\ttP=1-\ttM$ must be close to $1-\rhoP/\rho$ to reduce the amplitude of $\ZDsei$ such that our theory agrees with experimental observations.
Note that this value is close to unity in lithium-ion electrolytes.
Large cation transference numbers have been observed in mesoporous systems which immobilize anions \cite{Popovic2016}.
A similar situation could emerge in nano-sized SEI pores.
In principle, we could use other parameters such as the thermodynamic coefficient in the SEI to reduce the amplitude of $\ZDsei$.
However, this leads to unreasonable parameter choices because of the quadratic appearance of $\ttM$ in \cref{eq:rd}. 

In summary, large transference numbers $\ttP\approx1$ are necessary to avoid contradicting experimental observations.
Therefore, charge transport in the SEI has ``solid electrolyte character'' even if we assume ion transport in a liquid pore space.

\fsubsection[sec:resimp:equiv]{Summary: Equivalent Circuits}
\begin{figure}
\subfloat[\label{fig:equiva}]{}
\subfloat[\label{fig:equivb}]{}
\subfloat[\label{fig:equivc}]{}
\subfloat[\label{fig:equivd}]{}
\includegraphics[width=\columnwidth]{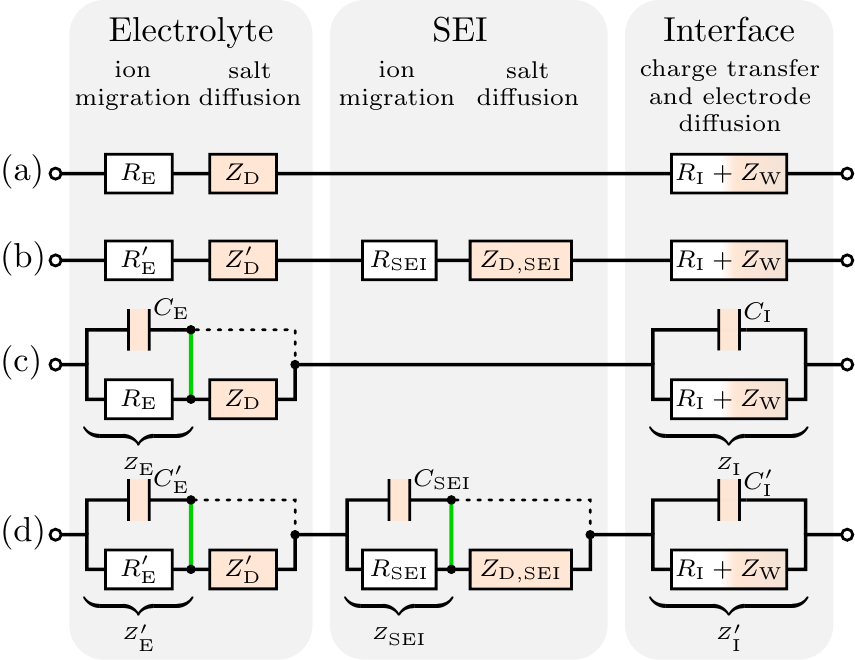}
\caption{\label{fig:equiv}
Summary of equivalent circuits.
Neutral model (a), neutral model with SEI (b), non-neutral model (c), and non-neutral model with SEI (d).
Circuit components with a frequency dependent resistance such as capacitors and Warburg elements are colored.
Here, $\ZD$ and $\ZDsei$ are Warburg Short elements.
Warburg Open elements $Z_\mathrm{W}$ describe diffusion processes in the electrode, see \cref{A-sec:calcS}.
Black dashed lines represent the technically correct circuit, the green \ins{connections are} approximations which we discuss in \cref{sec:resimp:equiv}.
}\end{figure}

We now transcribe \cref{eq:imp:zneutral,eq:resimp:ZneutralSEI,eq:znoseiimpfull,eq:zimpfull} into equivalent circuits.
These circuits are summarized in \cref{fig:equiv}. %a,fig:equivb,fig:equivc,fig:equivd}.
Our neutral models are equivalent to the circuits shown in \cref{fig:equiva,fig:equivb}.
This is different for non-neutral models.
The equivalent circuits shown in \cref{fig:equivc,fig:equivd} are an excellent approximation of the corresponding models provided the following conditions are met.

\textit{Firstly}, our model demands that the SEI phase and the electrolyte phase are large compared to the corresponding double layer thickness, i.e., $\Lsei\gg\lambdaDL'$, and $\Lely\gg\lambdaDL$.
This guarantees that diffuse layers do not overlap.
We assume this in our interface reaction model.

\textit{Secondly}, the interface reaction can only be represented as a standard RC element if the double layer width is constant in this frequency range.
This width is given by the inverse value of $\kP$ or $\tkP$.
These quantities become constant below the transition frequency $f_\mathrm{trans}$, see \cref{fig:kvsfreq}.
This transition coincides with the resonance frequency of the electrolyte semicircle.
Therefore, our analysis requires $\fint\ll\fE$ or $\RI\gg\lambdaDL/\conduc$.

\Cref{eq:znoseiimpfull,eq:zimpfull} are based on the green wiring in \cref{fig:equivc,fig:equivd}.
These approximations are valid if certain resonance frequencies are well separated, i.e., $\fD\ll\fE$, $\fDp\ll\fEp=\fE$, or $\fDsei\ll\fsei$.
This assumption separates the impedance into several distinct contributions. 
The dotted alternatives do not require these assumption, but they result in a single convoluted impedance expression instead.

\fsection[sec:resimp:exp]{Validation with Experiments}
\begin{figure*}[t]
\includegraphics{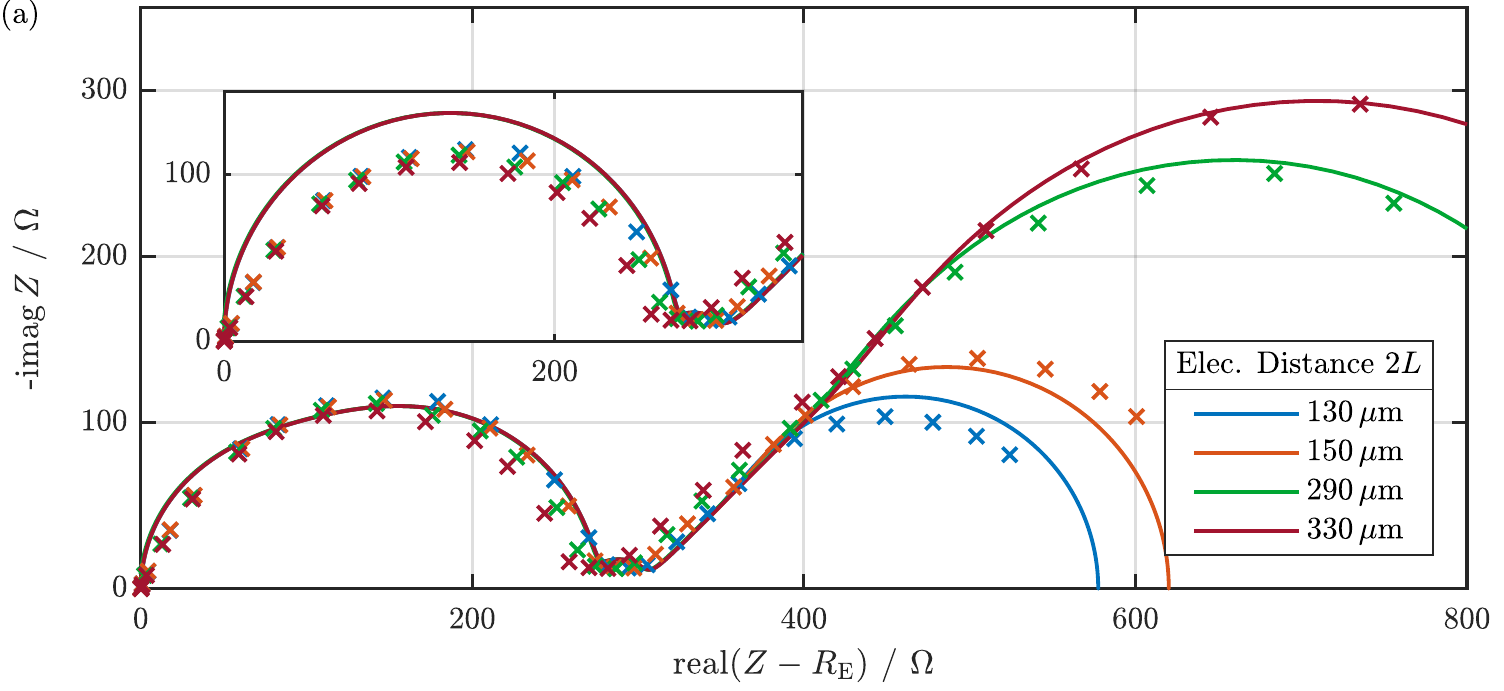}
\includegraphics{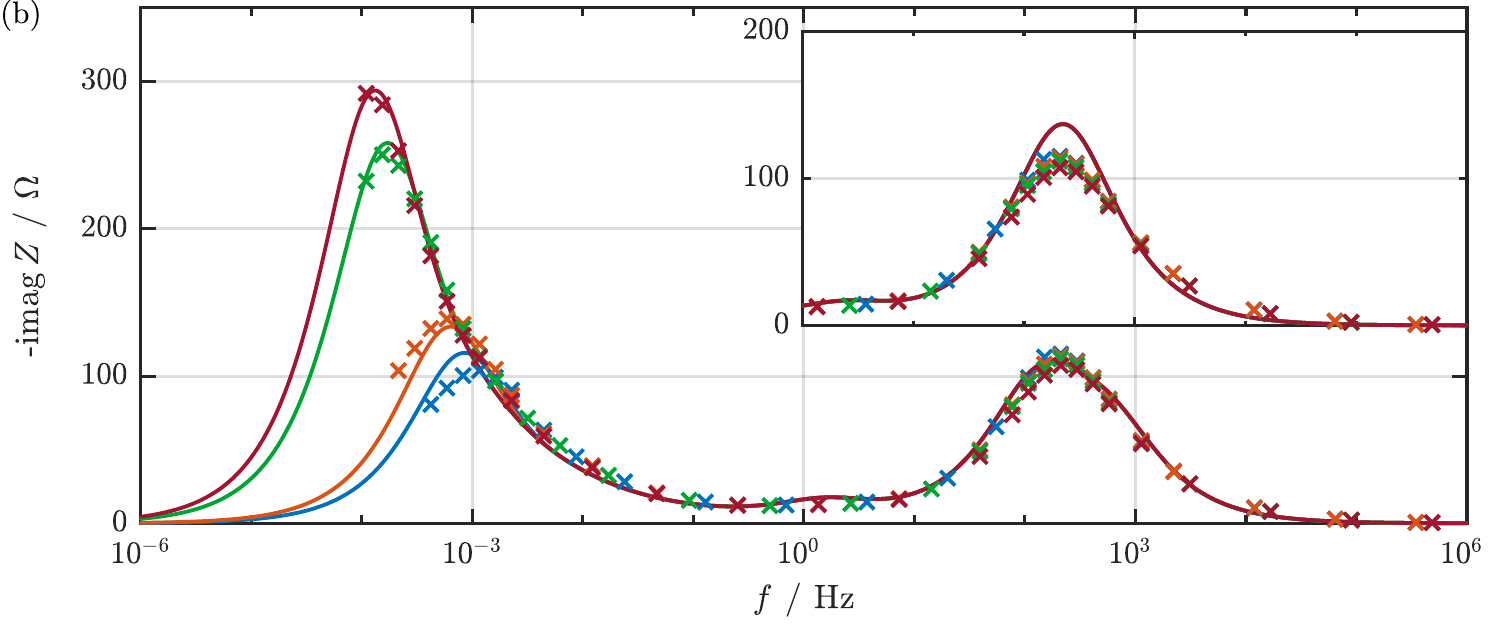}
\caption{\label{fig:impg4}
(a) Nyquist plot. (b) Bode plot, the $x$-axes of the inset aligns with the main $x$-axes.
Both plots show the impedance of a symmetric Li-Li cell with a Li-TFSI tetraglyme electrolyte.
Crosses mark experimental data points measured by Wohde et al. \cite{Wohde2016}.
Solid lines show our simplified impedance model according to \cref{eq:zimpfull}.
The main sets show this model with parameter set I whereas parameter set II is used in the insets.
Both parameter sets are listed in \cref{A-tab:param,A-tab:param2}.
}\end{figure*}
We now compare our impedance model to the experiment performed by F. Wohde et al. \cite{Wohde2016}
They measured the impedance of a symmetric Li-metal cell with planar electrodes.
The measurements were performed in a custom cell that allowed varying the distance between the electrodes.
After flooding the cell, its impedance was measured continuously in the high to intermediate frequency range.
In this way, the interface resistance was probed during the initial formation of the SEI.
It became constant after $24-48$ hours, indicating that stable surface films had formed.
Impedance measurements were performed only after this time.

We use the impedance data for the Li-TFSI electrolyte in a tetraglyme (G4) solution.
The Li-TFSI concentration of the electrolyte was equal to $\csalt=2.75\,$mol$\,$l$\inv$.
Measurements were performed for electrode distances of $130$, $150$, $290$, and $330\,\mu$m.
The impedance data is shown in \cref{fig:impg4}.
At first glance, four distinct features can be identified:

Firstly, the ohmic resistance of the electrolyte $\RE$ is equal to $7.5$, $8.8$, $17.6$, and $19.3\,\Omega$, increasing in proportion to the distance between the electrodes.
This contrition is subtracted from the data shown in \cref{fig:impg4}.
In this way, all remaining features align in the figure.

Secondly, the high frequency semicircle observed in these measurements is not influenced by the electrode distance.
Therefore, only interfacial processes like charge transfer reaction $\Zint$ or the SEI resistance $\Zsei$ can be assigned to this resonance.
A closer investigation reveals that this semicircle is depressed, suggesting that  it contains at least two of these resonances which overlap in frequency space.
These processes occur between 30 and 3000$\,$Hz and contribute approximately $250\,\Omega$.

Thirdly, a small impedance contribution of approximately 30$\,\Omega$ is present in the intermediate frequency range between 10 and 0.1$\,$Hz.
This resonance slightly overlaps with other resonances such that we cannot distinguish whether it has a semicircle or a Warburg type shape.
Due to its low frequency, only $\Zint$ and $\ZDsei$ are reasonable candidates.
We conclude this because the amplitude of this resonance does not scale with the electrode distance.
Additionally, typical resonance frequencies of the SEI are much larger, such that only these two candidates remain.

Fourthly, the measurements contain a diffusion type impedance contribution at very low frequencies ($10^{-4}-10^{-3}\,$Hz).
Its amplitude scales linearly with the electrode distance so that it can be identified as $\ZD$.

In conclusion, $\ZD$ can be directly identified in these measurements.
Assigning $\Zsei$, $\Zint$, and $\ZDsei$ to the features observed in the experiment is not as simple.
We suggest two different parametrizations of the impedance model for this reason.
\Cref{fig:impg4} shows the main parametrization in the outer plot and the alternative parametrization in the insets.
Let us first discuss parameters that these two options share.

The impedance model itself is over-parametrized.
This means that an impedance measurements alone is insufficient to determine each parameter of the model.
Instead, a subset of parameters must be known from independent experiments so that the remaining parameters can be identified.
Not all parameters are experimentally accessible.
For instance, the porosity $\porosity$ and tortuosity $\tortuosity$ of the SEI.
These parameters are chosen as $\porosity = 0.1$ and $\tortuosity=3450$.
Note that these parameters and the SEI thickness $\Lsei$ determine $\Rsei$, the resistance of the SEI see \cref{eq:crsei}.
Then, according to \cref{eq:resimp:fsei}, the only parameter remaining to tune the corresponding resonance frequency is $\tepsr$.

The resistance of the interface reaction is defined by the parameter $\RI$.
Originally, the model predicts that the corresponding resonance frequency is determined by $\RI$ and $\lambdaDL$ which depends on $\thermoSalt$, $\epsr$ and $\thermosplit$, see \cref{eq:resimp:fi,A-eq:resimp:lambdaDl}.
As discussed in \cref{sec:resimp:modelcorr}, model simplifications can lead to an incorrect prediction of the interface capacitance in real systems.
We consider this by adjusting the double layer thickness with the dimensionless parameter $\zeta$, see \cref{eq:resimp:lambdaDLcorr}.
In principle, $\lambdaDL$ could also be adjusted with other model parameters such as $\thermoSalt$, see \cref{A-eq:resimp:lambdaDl}.
However, this would lead to inconsistencies with other impedance contributions.

As mentioned above, we identify the diffusion impedance $\ZDp$ clearly in the experiment.
The resonance frequency of this effect is equal to $1.16$, $0.59$, $0.15$, and $0.11\,$mHz for the four different electrode distances.
From this we obtain $\Dsalt^*$ by inverting \cref{eq:resimp:fcritdiff} and averaging the four results, resulting in $\Dsalt^*=8.8^{-12}\,$m$^2\,$s$\inv$.
We estimate the partial molar volumes of the electrolyte with literature data for a different salt concentration \cite{Brouillette1998}.
$\RD$, the amplitude of $\ZD$ is then used to identify another parameter, see \cref{eq:rd}.
At this point, only the thermodynamic coefficient $\thermoSalt$ and the transference number $\tP$ are unknown.
Typically, $\thermoSalt$ can be measured independently, e.g., in a concentration cell experiment.
However, data for $\thermoSalt$ is not available in literature for the highly concentrated solution at hand.
We use a different approach for this reason.
%After fixing the partial molar volume, only $\tM$ or $\thermoSalt$ are left to determine $\RD$.
The transference number is taken from \cite{Wohde2016,dong2018efficient} which is calculated as follows
\begin{align}
\label{eq:resimp:tpapp}
\tP = \frac{\RE}{\RE+\RD} = 0.0245.
\end{align}
As discussed by Doyle and Newman \cite{Doyle1995}, this equation correctly gives the transference number for a dilute/ideal electrolyte.
Note that neither of these assumptions are applicable to this system.
%Therefore, this transference number can be considered as a rough estimate.
After choosing $\tP$, we calculate the thermodynamic factor with the impedance data resulting in $\thermoSalt = 6.2$.
The large thermodynamic factor illustrates the non-ideal behavior of the system.
This implies that \cref{eq:resimp:tpapp} gives a flawed estimate for the transference number.
However, this uncertainty could be removed by measuring the thermodynamic factor independently\cite{Ehrl2016} and using the impedance measurement to determine the transference number.

We now elaborate on the differences between both parameter sets.
They are summarized in the \ESI, see \cref{A-tab:param,A-tab:param2}\ins{\cite{Naejus1997}}.
In our \textit{main} parametrization, both $\Zsei$ and $\Zint$ overlap and form the high frequency semicircle.
$\Rsei$ and $\Rint$, the resistances of the SEI and the interface reaction are determined by choosing $\Lsei$ and $\RI$.
With the choice listed in \cref{A-tab:param}, these contributions amount to $\Rsei= 102\,\Omega$ and $\Rint =168\,\Omega$.
The resonance frequencies are then adjusted with the choices $\zeta=5$ and $\tepsr=131$.
Then, $\ZDsei$ must be assigned to the intermediate resonance.
At this point, $\ttP$ is the only parameter left to scale this impedance contribution.
The model fits well to the data if we choose $\ttP=0.9$.

In the \textit{alternative} parametrization, we attribute the high frequency resonance to $\Zsei$ alone.
To this aim, we increase $\Rsei$ by increasing the SEI thickness to $67\,$nm.
Then, the SEI resistance accounts for $\Rsei = 273\,\Omega$.
At the same time we adjust the relative permittivity of the SEI to move this semicircle to the correct resonance frequency, $\tepsr=347$.
$\Zint$ is attributed to the intermediate frequency resonance resulting in a lower value for $\RI$ so that $\Rint=23\,\Omega$.
However, shifting the corresponding resonance frequency to the intermediate regime requires a significant correction of the interface capacity, i.e., $\zeta=0.02$.
Finally, $\ZDsei$ is parametrized so that its amplitude is small. % i.e. by choosing $\ttM\approx\rhoP/\rho = 0.0134$, see \cref{eq:resimp:zwbsei2}.
In this case, $\ZDsei$ can be neglected if $\ttP=0.97$.

Both models show good qualitative agreement with the experiment, see \cref{fig:impg4}.
Naturally, the first parametrization has a better quantitative agreement with the data as it correctly predicts the depressed semicircle.
However, this depression could also be caused by an inhomogeneous distribution of SEI properties.
The main parametrization also requires more reasonable corrections.
For instance, the interface capacity needs to be corrected by a factor of $5$ in the main whereas a factor of $50$ is needed in the alternative one.
Both parametrization require large values for the relative permittivity of the SEI.
However, the value needed in the alternative option is even larger so that it is difficult to justify (135 vs. 347). %Additionally, although large, the relative permitivitty of the first parameterization is much smaller than in the second one (135 vs 347).
A better understanding of which parametrization is correct can be obtained by observing the high frequency impedance during the initial SEI formation.
This would easily allow the identification of SEI impedance contributions which increase during this time.

Both parametrization share a large lithium transference number to describe the transport in the SEI pores.
Here, the values $0.9$ and  $0.97$ are used, indicating that transport in the SEI pores is qualitatively different from transport in the bulk electrolyte.
As argued above, this behavior can be explained by immobilization of anions in the porous SEI structure \cite{Popovic2016}.
Although these two values are similar, they cause a significant change in the cell impedance.
In the first parametrization, $\RDsei$ is equal to $15\,\Omega$ creating a visible resonance between the interface semicircle and the low frequency finite-length Warburg.
In contrast, in the second parametrization, $\RDsei$ is equal to $1.5\,\Omega$ so that $\ZDsei$ is overshadowed and not visible.
Note that a smaller value of $\ttP$ would result in a much larger amplitude of $\ZDsei$ and can be ruled out.
Therefore, the model predicts lithium-ion transference numbers close to one in the SEI phase.

In conclusion, the assumption that charge transfer in the SEI is facilitated by liquid electrolyte in small SEI pores predicts the emergence of an additional feature in the cell impedance.
For the experiment at hand, this feature dominates the impedance response if the lithium-ion transference number in the SEI is not adjusted ($\ttP=0.0245 \rightarrow \RDsei = 1868\,\Omega$).
Our model only agrees with the experiment if large lithium-ion transference numbers are chosen in the SEI phase.
This indicates that lithium ions move in the solid phase of the SEI or that lithium-ion transport in the SEI pores has solid-electrolyte character.

\fsection[sec:summ]{Conclusions}
In this article, we present an analytical impedance model for a symmetric cell with planar electrodes.
We model the solid-electrolyte interphase (SEI) as a porous surface layer.
Our model relies on a thermodynamically consistent theory for electrolyte transport.
We take special care that transport parameters as the diffusion coefficient and the transference numbers are well defined.
This implies the consistent definition of a reference velocity, the center-of-mass velocity in our case.
Our analytic expressions for impedance spectroscopy show that this is especially relevant for concentrated electrolytes.

To reveal the parameter dependence of the impedance spectrum, we perform a step-by-step procedure.
We begin by calculating the impedance spectrum for locally electro-neutral systems without SEI. 
Finally, we relax the condition of local electro-neutrality and include transport through a porous SEI.
Most importantly, we describe diffusion impedances or Warburg short elements and reveal their dependence on the transference number.
Thus, we suggest to use impedance spectroscopy for the determination of the transference number or the thermodynamic factor in the bulk electrolyte and the SEI.

Our analytical impedance expressions are approximately described with common equivalent circuit models.
We identify and discuss the frequency range and parameter space in which equivalent circuit methods are valid.

We predict the thickness of charged electrochemical double-layers based on the Poisson equation.
Thus, in our model, the standard Debye length defines the double-layer capacitance and the resonance frequency of the interface reaction semicircle.
However, this frequency does not agree with experimental data for a Li-TFSI tetraglyme solution with Li-metal electrodes.
This indicates that charged double-layers of the reference state cannot be neglected when calculating the impedance response.
Furthermore, ion adsorption on the electrode surface contributes significantly to the interface capacitance of this system.

Our model explores the assumption that charge transport through the SEI is enabled by small pores which are filled with electrolyte.
This results in a second finite-length Warburg element in the impedance response of the cell.
Only a SEI specific Li$^+$ transference number close to one reduces the amplitude of this impedance contribution to levels that align with experiments. 
Therefore, charge transport in the SEI shows the characteristics of a solid electrolyte even if transport in a liquid environment is assumed.
We therefore propose to describe lithium-ion transport in the SEI with a specific theory for solid electrolytes.

\begin{acknowledgement}
This work is supported by the German Federal Ministry of Education and Research (BMBF) with the project Li-EcoSafe (03X4636A). The authors thank Prof. Bernhard Roling and Fabian S\"alzer for stimulating discussions.
\end{acknowledgement}

%% Bibliography
\bibliography{library}

\end{document}

%% file: head_packages.tex
\usepackage{amsmath}
\usepackage{textcomp}
\usepackage{graphicx}
\usepackage[T1]{fontenc}
\usepackage{enumerate}
\usepackage{array}
\usepackage[caption=false]{subfig}

\usepackage[nokeyprefix]{refstyle}
\usepackage{xr-hyper}
\usepackage{hyperref}
\usepackage{cleveref}

\usepackage{bm}
\usepackage{color}
\usepackage{lipsum}     % one column equation
\usepackage{tipa}       % textepsilon
\usepackage{units}      % \nicefrac
\usepackage{mathtools}
\usepackage{amssymb}    % \gtapprox

\usepackage{soul}
\usepackage{color} 

\newcolumntype{L}[1]{>{\raggedright     \let\newline\\\arraybackslash\hspace{0pt}}m{#1}}
\newcolumntype{C}[1]{>{\centering       \let\newline\\\arraybackslash\hspace{0pt}}m{#1}}
\newcolumntype{R}[1]{>{\raggedleft      \let\newline\\\arraybackslash\hspace{0pt}}m{#1}}
\newcolumntype{N}{@{}m{0pt}@{}}

\setcitestyle{super} % super script citations

\usepackage{longtable}
\usepackage{setspace}
\usepackage{etoolbox}
\AtBeginEnvironment{longtable}{\scriptsize}%singlespacing}

\usepackage{graphicx}

%% file: head_newcommands.tex
\newcommand{\titletext}{Theory of Impedance Spectroscopy for Lithium Batteries}

\newcommand{\ESI}{SI}

\newcommand{\ins}[1]{#1}
\newcommand{\remove}[1]{}

%% file: head_newcommands_basic.tex
\newcommand{\indexO}{1}      % Neutral, solvent
\newcommand{\indexT}{2}      % Neutral, solvent
\newcommand{\indexN}{{\mathrm{N}}}      % Neutral, solvent
\newcommand{\indexP}{{\mathrm{+}}}      % positive, cations
\newcommand{\indexM}{{\mathrm{-}}}      % negative, anions
\newcommand{\indexPM}{{\pm}}      % negative, anions
\newcommand{\indexA}{{\alpha}}
\newcommand{\indexB}{{\beta}}

\newcommand{\indexEQ}{{0}}

\newcommand{\indexI}{\mathrm{I}}        % Interface
\newcommand{\indexS}{\mathrm{S}}        % Solid, electrode
\newcommand{\indexE}{\mathrm{E}}        % Electrolyte
\newcommand{\indexBulk}{\mathrm{bulk}}  % Bulk
\newcommand{\indexSalt}{\mathrm{salt}}  % Bulk
\newcommand{\indexlin}{\mathrm{lin}}	% Symmetry soluation
\newcommand{\indexSEI}{\mathrm{SEI}}	% Symmetry soluation
\newcommand{\Dsymbol}{D}

\newcommand{\DAB}{D_{\indexA\indexB}^*}

\newcommand{\DAPcom}{D_{\indexA\indexP}}
\newcommand{\DAMcom}{D_{\indexA\indexM}}

\newcommand{\fluxN}{\text{N}}
\newcommand{\Nsymbol}{\fluxN}

\newcommand{\NP}{\fluxN_\indexP}
\newcommand{\NM}{\fluxN_\indexM}
\newcommand{\NA}{\fluxN_\indexA}

\newcommand{\csymbol}{c}
\newcommand{\cS}{{\csymbol}_\indexS}

\newcommand{\cAEQ}{{\csymbol}_{\indexA,\indexEQ}}

\newcommand{\cPMEQ}{{\csymbol}_{\indexPM,\indexEQ}}
\newcommand{\csaltEQ}{{\csymbol}_{\indexSalt,\indexEQ}}
\newcommand{\cN}{{\csymbol}_\indexN}
\newcommand{\cP}{{\csymbol}_\indexP}
\newcommand{\csalt}{{\csymbol}_\indexSalt}
\newcommand{\cPM}{{\csymbol}_\indexPM}

\newcommand{\cM}{{\csymbol}_\indexM}
\newcommand{\cA}{{\csymbol}_\indexA}
\newcommand{\cB}{{\csymbol}_\indexB}

\newcommand{\Phisymbol}{\phi}
\newcommand{\PhiS}{\Phisymbol_\indexS}

\newcommand{\PhiE}{\Phisymbol_\indexE}
\newcommand{\tPhiE}{\hat{\Phisymbol}_\indexE}
\newcommand{\PhiEEQ}{\Phisymbol_{\indexE,\indexEQ}}

\newcommand{\PhiI}{\eta}

\newcommand{\tcM}{\hat{\csymbol}_{\indexM}}

\newcommand{\vPhi}{\varphi}
\newcommand{\tvPhi}{\tilde{\varphi}}
\newcommand{\thermosplit}{\gamma}
\newcommand{\thermofac}{\mathcal{F}}
\newcommand{\thermoSalt}{\mathcal{F}_\indexSalt}
\newcommand{\thermoP}{\mathcal{F}_\indexP}

\newcommand{\cond}{\kappa}
\newcommand{\conduc}{\cond}

\newcommand{\zP}{z_\indexP}
\newcommand{\zM}{z_\indexM}

\newcommand{\zA}{z_\indexA}
\newcommand{\zB}{z_\indexB}

\newcommand{\nP}{n_\indexP}
\newcommand{\nM}{n_\indexM}
\newcommand{\nPM}{n_\pm}

\newcommand{\vsymbol}{\nu}

\newcommand{\vN}{{\vsymbol}_\indexN}

\newcommand{\vSalt}{{\vsymbol}_\indexSalt}

\newcommand{\vA}{{\vsymbol}_\indexA}

\newcommand{\fsymbol}{f}

\newcommand{\fA}{{\fsymbol}_\indexA}

\newcommand{\MN}{M_\indexN}
\newcommand{\MP}{M_\indexP}

\newcommand{\MA}{M_\indexA}

\newcommand{\tsymbol}{t}
\newcommand{\tP}{\tsymbol_\indexP}
\newcommand{\tM}{\tsymbol_\indexM}
\newcommand{\tA}{\tsymbol_\indexA}

\newcommand{\matD}{D}

\newcommand{\matT}{\bm{\mathcal{T}}}
\newcommand{\matI}{\bm{\mathcal{I}}}
\newcommand{\matA}{\bm{\mathcal{A}}}

\newcommand{\Dsalt}{D_\indexSalt}

\newcommand{\pxpy}[2]{\frac{\partial#1}{\partial#2}}

\newcommand{\dxdy}[2]{\frac{d#1}{d#2}}

\newcommand{\divv}{\mathbf{\nabla}\cdot}

\newcommand{\gradv}{\mathbf{\nabla}}

\newcommand{\divgradv}{\bm{\Delta}}

\newcommand{\partialt}{\partial_t}
\newcommand{\partialx}{\partial_x}

\newcommand{\inv}{^{-1}}

\newcommand{\boldcal}[1]{\bm{\mathcal{#1}}}

\newcommand{\vv}{\text{v}}
\newcommand{\vvoff}{\vv_\mathrm{off}}

\newcommand{\rhoP}{\rho_{\indexP}}

\newcommand{\rhoN}{\rho_{\indexN}}
\newcommand{\rhohatA}{\tilde{\rho}_{\indexA}}
\newcommand{\rhohatN}{\tilde{\rho}_{\indexN}}

\newcommand{\rhohatSalt}{\tilde{\rho}_{\indexSalt}}

\newcommand{\Jcom}{\mathcal{J}}

\newcommand{\muP}{\mu_\indexP}

\newcommand{\muN}{\mu_\indexN}
\newcommand{\musalt}{\mu_\indexSalt}
\newcommand{\mutsalt}{\tilde{\mu}_\indexSalt}
\newcommand{\muA}{\mu_\indexA}

\newcommand{\mutP}{\tilde{\mu}_{\indexP}}

\newcommand{\epsz}{\text{\textepsilon}_{0}}
\newcommand{\epsr}{\text{\textepsilon}_{\mathrm{R}}}
\newcommand{\tepsr}{\hat{\text{\textepsilon}}_\mathrm{R}}
\newcommand{\epszr}{\epsz\epsr}
\newcommand{\tepszr}{\epsz\tepsr}

\newcommand{\jint}{j_\indexI}
\newcommand{\jdls}{j_\indexS}

\newcommand{\Pflux}[1]{{\Nsymbol}_{#1}}
\newcommand{\porosity}{\varepsilon}
\newcommand{\tortuosity}{\tau}

\newcommand{\vecdeltac}{\vec{\delta\csymbol}}
\newcommand{\vecdeltatc}{\vec{\delta\hat{\csymbol}}}
\newcommand{\vecdeltacom}{\vec{\delta\csymbol}}

\newcommand{\deltacA}{\delta\cA}
\newcommand{\deltacB}{\delta\cB}
\newcommand{\deltacP}{\delta\cP}
\newcommand{\deltacM}{\delta\cM}
\newcommand{\deltatcM}{\delta\tcM}
\newcommand{\deltacS}{\delta\cS}

\newcommand{\deltaPhiE}{\delta\PhiE}
\newcommand{\deltavPhi}{\delta\vPhi}
\newcommand{\deltatvPhi}{\delta\tvPhi}
\newcommand{\deltaPhiS}{\delta\Phisymbol_{\indexS}}
\newcommand{\deltaPhiEom}{\delta\Phisymbol_{\indexE}}

%% file: head_newcommands_imp.tex
\newcommand{\tL}{L'}
\newcommand{\Lely}{\tL}%{L_\mathrm{SEI}}}
\newcommand{\Lsei}{\hat{L}}%{L_\mathrm{SEI}}}
\newcommand{\Ltot}{L}

\newcommand{\Qdls}{Q_\indexS}
\newcommand{\lambdaD}{\lambda_\mathrm{D}}
\newcommand{\lambdaDL}{\lambda_\mathrm{DL}}
\newcommand{\tlambdaDL}{\hat{\lambda}_\mathrm{DL}}

\newcommand{\N}{\mathcal{N}}
\newcommand{\transpose}{^\mathrm{T}}

\newcommand{\eigenvec}{\eta}
\newcommand{\teigenvec}{\hat{\eta}}

\newcommand{\veceta}{\vec{\eigenvec}}
\newcommand{\vecetaP}{\vec{\eigenvec}_\indexO}

\newcommand{\vecetaM}{\vec{\eigenvec}_\indexT}

\newcommand{\vecetaA}{\vec{\eigenvec}_\indexA}

\newcommand{\vectetaA}{\vec{\teigenvec}_\indexA}

\newcommand{\etaP}{{\eigenvec}_\indexO}
\newcommand{\etaM}{{\eigenvec}_\indexT}

\newcommand{\etaA}{{\eigenvec}_\indexA}
\newcommand{\etaB}{{\eigenvec}_\indexB}

\newcommand{\tetaA}{{\teigenvec}_\indexA}

\newcommand{\wavenum}{k}
\newcommand{\kP}{{\wavenum_\indexO}}
\newcommand{\kM}{{\wavenum_\indexT}}
\newcommand{\kA}{{\wavenum_\indexA}}
\newcommand{\kB}{{\wavenum_\indexB}}
\newcommand{\kAsq}{{\wavenum}_\indexA^2}
\newcommand{\ksei}{\hat{{\wavenum}}}
\newcommand{\tkP}{\hat{{\wavenum}}_\indexO}

\newcommand{\tkA}{\hat{{\wavenum}}_\indexA}

\newcommand{\iom}{i \omega}

\newcommand{\eiomt}{e^{\iom t}}
\newcommand{\emiomt}{e^{-\iom t}}
\newcommand{\eikx}{e^{ikx}}

\newcommand{\epikAx}{e^{ i \kA x}}
\newcommand{\emikAx}{e^{-i \kA x}}

\newcommand{\tDsalt}{\hat{D}_\indexSalt}
\newcommand{\tconduc}{\hat{\conduc}}
\newcommand{\ttP}{\hat{t}_\indexP}
\newcommand{\ttM}{\hat{t}_\indexM}

\newcommand{\coefficient}{C}

\newcommand{\CP}{\coefficient_{\indexO}}
\newcommand{\CM}{\coefficient_{\indexT}}
\newcommand{\CA}{\coefficient_{\indexA}}
\newcommand{\CAP}{\CA^+}
\newcommand{\CAM}{\CA^-}
\newcommand{\CApm}{\CA^\pm}
\newcommand{\CPpm}{\CP^\pm}
\newcommand{\CMpm}{\CM^\pm}

\newcommand{\tCP}{\hat{\coefficient}_\indexO}
\newcommand{\tCM}{\hat{\coefficient}_\indexT}

\newcommand{\vecC}{\vec{\coefficient}}

\newcommand{\coefficientp}{\Phi}
\newcommand{\PP}{\coefficientp'}

\newcommand{\Pp}{\coefficientp'}

\newcommand{\Pz}{\coefficientp}

\newcommand{\tPp}{\hat{\coefficientp}'}

\newcommand{\tPz}{\hat{\coefficientp}}

\newcommand{\potentialfactor}{\Pi}
\newcommand{\PiP}{\potentialfactor_1}
\newcommand{\PiM}{\potentialfactor_2}
\newcommand{\PiA}{\potentialfactor_\indexA}
\newcommand{\PiB}{\potentialfactor_\indexB}

\newcommand{\tPiA}{\hat{\potentialfactor}_\indexA}

\newcommand{\coefffunc}{\Gamma}
\newcommand{\GammaP}{\coefffunc_{1}}
\newcommand{\GammaM}{\coefffunc_{2}}
\newcommand{\GammaA}{\coefffunc_{\indexA}}

\newcommand{\GammaAp}{\coefffunc_{\indexA}'}
\newcommand{\GammaBp}{\coefffunc_{\indexB}'}

\newcommand{\tGammaA}{\hat{\coefffunc}_\indexA}

\newcommand{\RI}{\mathcal{R}}
\newcommand{\Rint}{R_\indexI}

\newcommand{\Zint}{Z_\indexI}
\newcommand{\Zintp}{Z_\indexI'}
\newcommand{\Cint}{C_\indexI}

\newcommand{\fint}{f_\indexI}
\newcommand{\fintp}{f_\indexI'}
\newcommand{\Rsei}{R_\indexSEI}
\newcommand{\Zsei}{Z_\indexSEI}
\newcommand{\Csei}{C_\indexSEI}
\newcommand{\fsei}{f_\indexSEI}

\newcommand{\fE}{f_\indexE}
\newcommand{\fEp}{f_\indexE'}
\newcommand{\fD}{f_\mathrm{D}}
\newcommand{\fDp}{f_\mathrm{D}'}
\newcommand{\RE}{R_\indexE}
\newcommand{\CE}{C_\indexE}

\newcommand{\ZE}{Z_\indexE}
\newcommand{\REp}{R_\indexE'}
\newcommand{\ZEp}{Z_\indexE'}
\newcommand{\tRE}{R'_\indexE}

\newcommand{\RSEI}{R_{\indexSEI}}
\newcommand{\RD}{R_\mathrm{\Dsymbol}}
\newcommand{\ZD}{Z_\mathrm{\Dsymbol}}
\newcommand{\RDp}{R_\mathrm{\Dsymbol}'}
\newcommand{\ZDp}{Z_\mathrm{\Dsymbol}'}

\newcommand{\tZD}{Z'_\mathrm{\Dsymbol}}
\newcommand{\RDsei}{R_{\mathrm{D},\indexSEI}}
\newcommand{\ZDsei}{Z_{\mathrm{D},\indexSEI}}

\newcommand{\fDsei}{f_{\mathrm{D},\indexSEI}}

%% file: head_modular.tex
\newcommand{\fsection}[2][]{\section{#2}\label{#1}}
\newcommand{\fsubsection}[2][]{\subsection{#2}\label{#1}}
\newcommand{\fsubsubsection}[2][]{\subsubsection{#2}\label{#1}}
\newcommand{\fsubsubsubsection}[1]{\textit{#1:}}

\crefname{equation}{eq.}{eqs.}
\Crefname{equation}{Equation}{Equations}
\crefname{figure}{fig.}{figs.}
\Crefname{figure}{Figure}{Figures}
\crefname{section}{sec.}{secs.}
\Crefname{section}{Section}{Sections}

\newcommand{\fwidebegin}{\begin{strip}}\newcommand{\fwideend}{\end{strip}}\SectionNumbersOn\usepackage{cuted}	% widetext alternative strip works only with achemso
\usepackage[sort&compress,numbers,super]{natbib}
\setkeys{acs}{doi = true}

\hypersetup{pdftitle=\titletext,pdfkeywords={Impedance,SEI,Li-ion,Battery},pdfauthor={Fabian Single, Birger Horstmann, Arnulf Latz}}